\documentclass[10pt,aps,pbl,keywords,groupedaddress,nofootinbib]{revtex4}
\usepackage{graphicx}
\usepackage{color}

\begin{document}

\title{Monte Carlo Simulations of DNA Damage and Cellular Response to Hadron Irradiation}

\author{M. Loan$^{a,b}$\footnote{Corresponding author}, B. Freeman$^{c}$, A. Bhat$^{d}$, M.
Tantary$^{e}$, M. Brown$^{d}$ and K. Virk$^{c}$\\
(KCST-CVMC Collaboration)}
\affiliation{$^{a}$Department of Physics, Kuwait College of Science and Technology, 28007, Kuwait \\
$^{b}$ANUC, Australian National University, Canberra, 2000,
Australia\\
$^{c}$Department of Computer Science, Kuwait College of Science and Technology, 28007, Kuwait\\
$^{d}$Department of Oncology, Clinch Valley Medical Center, Richlands, Virginia, 24641, USA\\
$^{e}$Department of Internal Medicine, Clinch Valley Medical
Center, Richlands, Virginia, 24641, USA\\
$^{c}$Department of Computer Science, Kuwait College of Science
and Technology, 28007, Kuwait}
\date{\today}

\begin{abstract}

Numerical simulations are performed on a stochastic model based on
Monte Carlo damage simulation process and Markov Chain Monte Carlo
techniques to investigate the formation and evaluation of isolated
and multiple DNA damage and cellular survival by light ionizing
radiation in a colony of tumour cells. The contribution of the
local clustering of the strand breaks and base damage is taken
into account while considering double-strand breaks (DSBs) as
primary lesions in the DNA of the cell nucleus induced by ionizing
radiation. The model incorporates the combined effects of
biological processes such as the tumour oxygenation, cellular
multiplication and mutation through various probability
distributions in a full Monte Carlo simulation of fractionated
hadrontherapy. Our results indicate that the linear and quadratic
parameters of the model show a negative correlation, for protons
and helium ions, which might suggest an underlying biological
mechanism. Despite using a model with quite different descriptions
of linear and quadratic parameters, the observed results for
linear parameter show largely reasonable agreement while the
quadratic parameter consistent deviations from the results
obtained using the LEM model at low LET. In addition to the LET
dependence, RBE values showed a strong dependence on $\alpha
/\beta$ ratio and a considerable scatter for various particle
types indicate particle specific behaviour of initial its slope.
The surviving curves show a non-linear dose-response suggesting
that interaction among DSBs induced by ionizing radiation
contribute significantly to the quadratic term of the model.
Nonetheless, our simulation results suggest that not only is the
model suited for effectively predicting the relative biological
effectiveness of the charged particles at low LET and different
survival, but also to accuracy in prediction of
cell-killing in radiotherapy such as hadrontherapy.\\


\end{abstract}

\maketitle

\section{Introduction}

Hadron radiotherapy, one of the established modalities for the
treatment of cancer-based on irradiating tumours with accelerated
light ions is a promising method for the treatment of specific
types of cancer
\cite{Mario2017,Francesco2015,Regler2009,Pavel2006,Mayer2004,Krengli2004,Brahme2001,Tilly1999}.
Their dose-distribution characteristic culminating with higher
rate of energy loss at the end of their ranges, resulting in a
sharp decrease beyond the Bragg peak allows for a high conformity
to the deep-seated tumours while sparing healthy tissues
\cite{Tom2015,Linz2012,Durante2010,Amaldi2005}. With the range
modulating techniques to spread out the Bragg peak to cover the
whole target, the ionized dose is delivered with a spectrum of
different linear energy transfers (LET) at each point in the
spread-out Bragg peak. Since the high-LET component has an
increased biological efficiency resulting in the induction of
enhanced and unrepairable biological damage, quantified in terms
of Relative Biological Effectiveness (RBE), the ionized particles
show enhanced biological effectiveness in eradicating the tumour
cells
\cite{Paganetti2013,Girdhani2013,Friedrich2012a,Friedrich2012,Paganetti2002}.
Apart from protons, the most frequently used ion species is
carbon; however, heavier ions such as oxygen are available in some
centers too, and irradiation with lighter ions such as helium or
lithium may be advantageous for some tumours.

The induction of radiation-induced DNA breaks is of central
importance of radiation-induced cell death. The reproduction of
the radiation-induced yield of single- and double-strand breaks is
taken as a benchmark for numerical simulation models for the
estimation of the radiation damage to DNA. Differences in the
biological response induced by ionized particles protons compared
to photons reported in last few years raise the concerns regarding
the analysis of DNA damage induction and processing. Besides
producing approximately 20-25 times more single-strand breaks
(SSBs) than double-strand breaks (DSBs) per grey and a base damage
of 2,500-25,000 per grey per cell in a typical mammalian cell
\cite{Ward1985,Ward1988,Ward1995}, low-LET ionized radiations
effectively produce multiple damage sites (MDS) consisting of two
more strand breaks or base damage within a few turns of DNA helix.
These strand breaks are detected as an SSB in most experimental
studies. Whereas, DSBs and MDS act as the primary cause of cell
kill by the radiation.

DNA isolated and multiple strand break inductions by different
radiation qualities show that clear differences emerge when
looking at the rejoining process
\cite{Belli2002,Leloup2005,Hada2006,Calugaru2011}. The lack of a a
general relationship between cellular effects with respect to
radiation quality has made the precise measurements of DSBs and
MDS with radiation quality. Earlier measurements of yields of
strand breaks showed approximately slight or no variation with
radiation quality. Some earlier studies reported the relative
biological effectiveness (RBE) for induction of DSBs by the
protons relative to 200-250 keV/$\mu$m x-rays radiations vary
between 0.8 and 1.1 using the sedimentation technique as well as
filter elution
\cite{Prise1998a,Prise2001,Belli1994,Frankenberg1999}. Data on RBE
for induction of DSBs in human cells using protons, helium and
carbon ions up to about 120 keV/$\mu$m were reported in the range
1.35 which is much lower compared to that for cell killing and
malignant cell transformation
\cite{Belli2002,Belli2001,Belli2000c,Pinto2002}. These data are in
contrast to RBE values of 2.6 for induction of DSBs found in yeast
using 120 keV/$\mu$m helium ions using neutral sedimentation.
These significant differences in RBEs for induction of SSBs and
DSBs, as well as the cellular effects, reflect different
reparability of strand breaks as a function of radiation quality.

The ideal use of hadron therapy heavily relies on modelling
\cite{Carante2015,Naqa2012,Wang2010,Dale2009}. Clear understanding
of physical and biological processes underlying the
hadro-biological the mechanism is essential for the optimal uses
of treatment planning. However, existing hadrontherapy treatment
planning approaches are largely based on the effectiveness of the
physical processes involved compared to the the biological
response elicited in cells and tissues by hadrons and ions. Most
of the treatment planning is based on interpolating the
linear-quadratic (LQ) parameters of LQ fits the experimental data
on surviving curves while assuming similarity in biological
effects among heavy ions species of similar LET values
\cite{Kramer2000a,Kramer2000b,Kagawa2002}. This indicates that the
biological processes of damage induction and repair have not been
considered explicitly and is reflected in the large uncertainties
that can be found in the literature regarding the $\alpha$,
$\beta$ parameters of the LQ model and in the definition of
RBE-LET relationship for light and heavy ions. More generally,
evidence has also been reported showing {\emph{ in vitro}}
measurement with V79 cells having a low $\alpha /\beta$ while
{\emph{in vivo}} experiments showed a high $\alpha /\beta$ ratio
\cite{Paganetti2013}. Since RBE has a dependence on this ratio, a
direct quantitative comparison between these measurements will be
misleading. The published data on RBE values also showed large
fluctuations in RBE data in tumour cells. as well as in
neighbouring healthy tissues \cite{Friedrich2012b}.

An alternative approach is to incorporate the biological
parameters into the ionized radiation response models. This has
gained considerable attention including numerous analytic and
numerical models which seek to include physical as well as
biological effects. A very significant improvement on the front
has been achieved in some most recent studies on track structures
and resulting DNA damage and repair processes through pathways,
suggesting a non-trivial relationship between radiation-induced
DSBs in the cell nucleus and their LET dependencies as well as the
on the probability of cell death induced by the DSB effects
\cite{Wang2018,Ramin2017,McMahon2017,Friedland2017,McMahon2016}.

The aim of this study is to build on and extend the previous
detailed description of recently published hadrontherapy models on
radiation-induced DNA strand breaks, DNA repair, and cellular
survival \cite{Wang2018,Friedland2017}. Whereas the mixed beam
models, such as the micro-dosimetric kinetic model (MKM)
\cite{Hawkin1998,Hawkin2003}, the local effect model (LEM)
\cite{Scholz1996a,Scholz1996b,Scholz1997}, and the mechanistic
models including repair-misrepair fixation (RMF) model
\cite{Karge2018} and BIANCA model have been successful in allowing
to predict the relationship between physical parameters of
radiation and cell survival, we use the mechanistic model proposed
by Wang {\emph{et al}., \cite{Wang2018} in this study as this
model offers predictions at the molecular and cellular levels that
are quantitively described by only two input parameters; the
average number of primary particles that cause DSB, and the
average number of DSBs yielded by each primary particle that
caused DSB. Furthermore, the model provides a simpler structure in
terms of the quantification of the effects of dependence of linear
and quadratic components on LET and dose and calculation of RBE
for the same type of cells exposed to different particle spices at
different LET.

Considering that DSBs are the initial lesion in the DNA of the
cell nucleus induced by ionizing radiation, compared to DSBs that
occur spontaneously during cellular processes at quite significant
frequencies, while taking into account the non-homologous
end-joining pathway as domain pathway of DSB repair in mammalian
cells, we simulate a modified model that incorporates biological
processes such as adaptive response and hypoxic effects with
sufficient number of randomly selected cells surpassing the
approximate avascular growth phase and approaching clinical
levels. The induction of radio-adaptive response in the model
signifies the ability of low dose radiation to induce cellular
changes that alter the level of subsequent radiation-induced or
spontaneous damage. These effects appear to be especially crucial
in the case of exposure to low doses and dose rates of ionizing
radiation
\cite{Sawant2001,Feinendegen96,Feinendegen99,Azzam96,Mitchel99,William2003,de2004,Shankar2006}.
Recent studies have shown that there is conclusive evidence for
some cells that radio-protection induction from an adaptive
response may be activated by a few or single charged particle
track traversals through the sensitive region of individual cells
\cite{Leonard2000,Leonard2005,Leonard2007a,Leonard2007b,Leonard2007c}.
Equally important is the incorporation of radiobiological
differences between acute and chronic hypoxic cells. The modified
model simulates chronically hypoxic cells that may have low or
depleted energy reserves which would impair their repair
mechanisms in contrast with acutely hypoxic cells that are repair
competent. Hypoxia is modelled through the oxygen level allocation
based on $pO_{2}$ probability distribution.

The improved model also attempts to address the discrepancies and
fluctuations in RBE observed in the results of different
experiments and the lack of complete inclusion of the underlying
biological processes with light ions, especially in the low LET
region, by attributing the roles played by biological parameters
such as tissue type, oxygenation level, and balance between an
increase in radiation induced DSB yield and increased loss of DNA
fragments induced along the track of primary particles. Monte
Carlo simulation in this LET region will be highly useful until a
more detailed knowledge of these effects is available.

The rest of the paper is organised as follows: A probabilistic
model, that incorporates the combined effects radiation-induced
isolated and multiple DNA strand breaking, adaptive response,
tumour oxygenation, cellular multiplication, and mutation in a
full Monte Carlo simulation of fractionated hydrotherapy is
outlined in Sec. II. The details of hybrid Monte Carlo techniques
to simulates a random cell number, sufficient enough to surpass
the approximate avascular exponential growth phase to reach a cell
colony at the clinical level, within a nonspatial environment, is
given in Sec. III. We present and discuss our results in Sec. IV.
Here, the improved probabilistic model results, with stochastic
biophysical input, are validated and compared with results
obtained by various other simulation and experimental studies. In
particular, a systematic comparison of linear and quadratic
parameter predictions of the improved model with those of LEM
model is of interest, since LEM model is already used by treatment
planning systems of European dual ion therapy facilities for
biological optimization in carbon ion therapy. Section V is
devoted to the summary and concluding remarks. We hope that the
quantitative approach illustrated here will facilitate
improvements to our understanding of physical and biological
processes, in hadrontherapy, that will no doubt become possible
and necessary as additional experimental evidence becomes
available.

\section{Improved Probabilistic Model with Stochastic Input}

\subsection{Cell Survival Model}

The probability distribution of cell death with radiation in a
fully oxic population can be described in LQ theory by an
equivalent equation:
\begin{displaymath}
P_{ox}(D) = 1-exp(-\alpha D-\beta D^{2})
\end{displaymath}

where the linear component $\alpha$ characterize a single lethal
event made by one-track action and the quadratic component
$\alpha_{2}$ characterizes the accumulation of sub-lethal events
made by two-track action for oxic conditions. However, both these
events describe the DNA double-strand breaks and do not take into
account the damage caused by the DNA single-strand breaks. Studies
have shown that single-strand damage caused by x-rays is far more
than the number of double-strand damage
\cite{Jioner2009,Philp2010,Reza2011}. Also, both the experimental
and model studies on isolated and multiple strand damage by the
ionizing radiation and repair process
\cite{Wang2018,Meylan2017,Friedland2017,McMahon2017,McMahon2016,Ramin2017,Ding2009},
have shown that the conventional linear-quadratic model is poorly
suited to understanding these factors, as its empirical parameters
are only indirectly linked to the mechanistic drivers of radiation
response, making it difficult to predict quantitatively the impact
of a given mutation. This is particularly true when multiple genes
are mutated, as occurs commonly in cancer.

To simulate the cell death probability due to charged particle
irradiation, we use the recently proposed mechanistic model that
incorporates the kinetics of cell death induced by the DSBs effect
and DNA repair through different pathways and cell death processes
\cite{Wang2018}. The model describes the yield of radiation
induced DSBs through contributions of interaction among DSBs
induced by different primary particles described by the average
number of primary particles which caused DSB and DSBs induced by
single primary particle and their interactions, including the
cluster DNA damage effect and the overkill effect by the average
number of DSBs yielded by each primary particle that caused DSB .

Assuming that the number of DSBs yielded by the primary particle
is Poisson-distributed \footnote{For light and heavy ions at high
LETs, the deviations of lethal lesions from Poisson distribution
are significant. It is argued that increasing LET causes deviation
from the Poisson distribution by non-random clustering of lethal
lesions in some cells}, the average number of primary particles
that cause DSBs is given by
\begin{displaymath}
n_{p} = \frac{YD}{\lambda}(1-e^{-\lambda})
\end{displaymath}
and the average number of DSBs yielded by each primary particle
that caused DSB is given by
\begin{displaymath}
\lambda_{p} = \frac{\lambda}{1-e^{-\lambda}},
\end{displaymath}
where $\lambda =(N/n)$ is the DSB yield per cell per primary
particle that depends on average number of radiation-induced DSBs
per cell, $N$ (=$Y\times D$) and number of primary particles, n
($\propto 1/LET$) passing through the nucleus.

The model considers the repair of DSBs through nonhomologous
end-joing (NHEJ) pathway as the dominating pathway in mammalian
cells and assigns the probability of a DSB being correctly
repaired as
\begin{displaymath}
P_{correct}= \mu_{x}P_{int}P_{track},
\end{displaymath}
where
\begin{displaymath}
P_{int}= \frac{1-e^{\eta (\lambda_{p})n_{p}}}{\eta (\lambda_{p})n_{p}}\\
P_{track}= \frac{1-e^{\xi \lambda_{p}}}{\xi \lambda_{p}}
\end{displaymath}
$\mu_{x}$ is the average probability of DSB end joining with the
other end from the same DSB correctly with $\mu_{y}$ as the
sensitivity of an error repair and where $\xi\lambda_{p}$ is the
average probability of a DSB end being joined with a DSB end from
a different DSB induced by the same primary particle. Finally, the
probability of a DSB contributing to cell death is given by
\begin{displaymath}
P_{contrib}= \frac{1-e^{\phi \lambda_{p}}}{\phi \lambda_{p}}
\end{displaymath}
The average number of lethal events, $N_{death}$, is then given by
\begin{displaymath}
N_{death}=\mu_{y}\times P_{contrib}\times (1-P_{correct})
\end{displaymath}
This results in the following probability model, with the similar
form as the LQ model, for the cell death \cite{Wang2018}:
\begin{equation}
P_{ox}(D)\propto \left[1-e^{-\alpha D-\beta D^{2}}\right],
\label{eqn1}
\end{equation}
where
\begin{eqnarray}
\alpha & = & Y\times
\left[\frac{1-e^{\phi\lambda_{p}}}{\phi\lambda_{p}}\right]\times
\left[1-\mu_{x}\left\{\frac{1-e^{\xi\lambda_{p}}}{\xi\lambda_{p}}\right\}\right]\nonumber\\
& & \times \mu_{y}\\
\beta & = &
\frac{1}{2}\eta(\lambda_{p})\frac{Y}{\lambda_{p}}\times Y\times
\left[\frac{1-e^{\phi\lambda_{p}}}{\phi\lambda_{p}}\right]\nonumber\\
& & \times
\left[1-\mu_{x}\left\{\frac{1-e^{\xi\lambda_{p}}}{\xi\lambda_{p}}\right\}\right]
\mu_{x}\mu_{y}. \label{eqn3}
\end{eqnarray}
are the improved model linear and quadratic parameters,
respectively. The functional dependance of $\alpha$ and $\beta$ on
the LET and dose $D$ can be seen through $\lambda$, $n_{p}$ and
$\lambda_{p}$. Since at low-LET, dose response of DSB induced by
radiation is linear, therefore the cell survival curve is in
agreement with the LQ model. However, for high-LET radiations,
when delivering the same dose to the nucleus as low-LET
radiations, the number of primary particles causing DSB is much
smaller, and the contribution of interaction among DSBs induced by
different primary particles to cell death would be vanishingly
small, therefore the cell survival curves tend to follow the the
exponential models \cite{Wang2018}. The dependence of alpha and
beta on the dose range has shown an impact on the alpha/beta ratio
determined from the survival data \cite{Garcia2006,Garcia2007}.
The low-dose region had a significant influence that could be a
result of a strong linear, rather than quadratic component,
hypersensitivity, and adaptive responses. Such a dependence serves
as a caution against using cell survival data for the
determination of $\alpha / \beta$) ratio.

The model is further improved by quantifying the biological effect
of oxygen concentrations on tumour cells in terms of the oxygen
enhancement ratio ($OER$). Assuming that (i) the damage response
of DNA single-stranded breaks and double-stranded breaks are two
relatively independent processes involving different signaling
pathways, (ii) oxygen distribution in tissue has a cylindrical
symmetry, and (iii) each hypoxic cell as well as each capillary
consumes oxygen spatially at an equal and constant rate (no
intravascular resistance), the improved cell death probability can
be modelled by the joint probability distribution
\begin{equation}
p_{hyp}(D)\propto \left[1-e^{-\alpha_{m}D\cdot OER(D,L,p)
-\beta_{m}D^{2}\cdot OER^{2}(D,L,p)}\right] \label{eqn4}
\end{equation}

The $OER$ is determined from dose-LET and $pO_{2}$ dependent model
\cite{Tatiana2011a}
\begin{equation}
OER(D_{h},L,p)= \frac{2D\alpha_{2}(L,p)}{\Phi
(L,p)-\alpha_{1}(L,P)}, \label{eqn5}
\end{equation}
where
\begin{displaymath}
\Phi (L,p)
=\sqrt{\rho^{2}_{1}(L,p)+4\rho_{2}(L,p)(\rho_{1}(L,p)D_{h}+\rho_{2}^{2}(L,p)D^{2}_{h})}.
\end{displaymath}
The dependence of $\rho_{i}$ on $pO_{2}$ and LET, in the
clinically relevant LET region, is given by
\begin{eqnarray}
\rho_{1} (L,p) & = & \frac{(a_{1}+a_{2}\cdot L)\cdot p +
(a_{3}+a_{4}\cdot L)\cdot K}{p+K} \nonumber\\
\rho_{2} (L,p) & = & \rho_{2}(p)=\frac{(b_{1}\cdot L)\cdot p +
b_{2}\cdot K}{p+K}, \nonumber
\end{eqnarray}

where $L$ is linear energy transfer, $a_{i}$ and $b_{i}$ are
constant coefficients \cite{Tatiana2011b} and $K$ represents the
oxygen concentration around $2.5 - 3$ mmHg. Eq. \ref{eqn5} can be
used to describe the $OER$ for various different radiation types
at low- and high-LET regions and for various oxygen levels
relevant to theoretical and clinical situations. We observe the
model predictions relative to the oxygen effect with light and
heavy ion irradiation and compare the results with preclinical and
clinical studies.

Equation (\ref{eqn4}) is randomly simulated to determine the
probability of cell death from its surviving probability. We will
examine and compare the effects on the slopes of the survival
curves for both well-oxygenated and hypoxic tumours for low- and
intermediate LET hadrons and charged ions using a hadron specific
approach. Since the $OER$ for protons is similar to x-rays
\cite{Tsujii2008,Tsujii2006,Suzuki2006}, namely $2.5$ to $3$, it
will be interesting to know how will the different fractionation
schedule affect $OER$. This is particularly important in light of
the ambiguous relationship between tumour oxygenation and the
model parameters, $\alpha$ and $\beta$.

\subsection{$\alpha /\beta$ ratio and $RBE$ characterization}
The ratio of intrinsic parameters, $\alpha$ and $\beta$, of the LQ
model, is a measure of the fractionation sensitivity of the cells.
The cells with a lower $\alpha /\beta$ are more sensitive to the
sparing effect of fractionation. The determination of
hadrotherapeutic outcome and therapeutic window strongly depends
on a reliable estimation of parameters $\alpha$, $\beta$ and
$\alpha /\beta$. Because of the mechanistic nature of the above
model, it can be directly extended to validate the DSB
distribution caused by x-rays and charged particles. Therefore,
the model is able to make predictions for $\alpha /\beta$ ratio
and RBE at a different survival with hadrons and heavy charged
ions. The modelled parameters $\alpha$ and $\beta$ for the same
type of cells irradiated by different radiation types at different
LET can be used to reflect on the ratio $\alpha /\beta$ as an
indicator of cellular repair capacity. Using Eqs. (2) and (3), the
ratio $\alpha /\beta$ is given by \cite{Wang2018}
\begin{equation}
\alpha /\beta =
\frac{1-\mu_{x}\bigg(\frac{1-e^{-\xi\lambda_{p}}}{\xi\lambda_{p}}\bigg)}{(1/2)\mu_{x}\eta(\lambda_{p})Y/\lambda_{p}},
\label{eqn6}
\end{equation}
where $\eta$ for a given track is determined by the distribution
of DSBs created by the track. The ratio is dominated by the
interaction of DSBs induced by different primary particles in the
limit $\lambda_{p}\rightarrow 1$ for photons and low-LET
irradiation, whereas for hadrons and ions at higher LET, the ratio
is attributed to $Y/\lambda_{p}$. Because RBE values involve two
parameters of LQ model, we input the role of improved $\alpha
/\beta$ ratio to obtain the explicit dependance of RBE on the
hadron dose as well as the LET from selected hadron:
\begin{eqnarray}
RBE(L,D,(\alpha /\beta)_{x}) & =&
\frac{1}{D_{p}}\left[\sqrt{\frac{1}{4}\bigg(\frac{\alpha}{\beta}\bigg)^{2}_{x}
+\Gamma}
-\frac{1}{2}\bigg(\frac{\alpha}{\beta}\bigg)_{x}\right],\nonumber\\
\end{eqnarray}
where
\begin{displaymath}
\Gamma = \bigg(\frac{\alpha}{\beta}\bigg)_{x}\frac{\alpha
(L)}{\alpha_{x}}D+ \frac{\beta (L)}{\beta_{x}}D^{2}.
\end{displaymath}
We compare the our Monte Carlo estimates for cell-survival
fraction to the experimental data to validate the accuracy of the
above RBE model.

\section{Monte Carlo Implementation}
\subsection{Hybrid Algorithm}

The numerical simulations were divided into two parts; the first
part of the hybrid algorithm uses Monte Carlo damage simulation
(MCDS) algorithm \cite{MCDScode} to simulate the formation of
isolated and multiple damaged DNA sites by radiation of various
species and different LETs. This algorithm captures the trend in
DNA damage spectrum with the possibility that the small-scale
spatial distribution of elementary damages is governed by
stochastic events and processes \cite{Stewart2015}. The use of
this quasi-phenomenological algorithm is to provides
nucleotide-level maps of the clustered DNA lesions, including
simple and complex forms of the single-strand break (SSB) and DSB
and to avoids the initial simulation of the chemical processes.
This algorithm also allowed a range of particle energies has been
expanded, and the induction of damage for arbitrary mixtures of
charged particles with the same or different kinetic energies can
be directly simulated. To examine the effects on damage complexity
of the direct and indirect mechanisms, a modified version of MCDS
algorithm was used to mimic reductions in the number of strand
breaks and base damages associated with exposure to an extrinsic
free radical scavenger, dimethyl sulfoxide, that offers protection
against both strand breakage and base damage.

The second part uses a stochastic Monte Carlo technique to
simulate the evaluation of cell survival. Accordingly, the hybrid
code is divided into two different parts of which each uses
different Monte Carlo algorithms. Radiation-induced DSB yield per
cell per Gy and DSB yield per cell per primary particle were
directly obtained with MCDS algorithm. These parameters were used
to estimate the initial slopes of linear and quadratic
coefficients for the later use in the stochastic Monte Carlo
algorithm for the evaluation of cell survival. The software
evolved from the MCDS code and many routines were adapted from an
existing stochastic serial code to return a large-scale framework
despite the different target theory. Each part of the hybrid code
is configured to output its own intermediate results so that code
can be validated against the results from various numerical and
experimental methods.

The stochastic Monte Carlo technique
\cite{Fornalski2011,Fornalski2014} is used to develop the dynamics
of cell death governed by Eq. (\ref{eqn4}) using a predefined
fractionation schedule of one irradiation per day. The Monte Carlo
algorithm employed consists of a nested numerical loop over
radiation dose, a loop over the time steps, a loop over oxygen
supply, a loop over the age of each initial cell and a loop over
the cells. At each step, the cells may exhibit the possible
altered states or division. The algorithm uses continuous and
differentiable probability distribution relationships to describe
the biophysical effects of cell colony. At each time step, a
stochastic tree of probabilities is applied to every $i$-th cell
which alters the state of the cell depending on whether it has
been irradiated or not. The modification in the probability tree
can be made easily and new branches can be added to incorporate
additional biological effects or modify ones detailing the
mechanism in cell eradication.

A regular grid of capillary cells is used to initiate the
oxygenation. To ensure a significant supply of oxygen for the
tumour cells, a regular grid of capillary cells is used. As the
tumour cell splits, the daughter cells are placed at the adjacent
positions extended in the direction of the tumour boundary. For an
occupied position, the respective cell is shifted to one of its
neighbouring positions. The shifts are repeated iteratively until
a free grid position is available. Our simulation results focus on
simulating oxic, moderately hypoxic and severely hypoxic tumours
while assigning uniform, log-normal and normal probability
distributions of $pO_{2}$ values based on published data
\cite{Forster17,Lartigau98} to allocate cellular oxygenation.
Compared to earlier models that simulate spatial oxygenation
distributions by assuming a spherical geometry and use the radial
distances of cells from the tumour periphery to determine
oxygenation \cite{Maseide00,McElwain93,Borkenstein04}, this
technique method of allocating oxygenation levels from a $pO_{2}$
probability distribution is simple and user-friendly and has the
advantages of flexibility as it easily allows for the $pO_{2}$
distributions and variation during the growth or treatment in a
single simulation, if required. It also enables easy expansion of
the model to more refined forms relationships between the oxygen
allocation and probability distributions during future studies
\cite{Forster17}.

We start the algorithm with a colony of $10^{6}$ cancer cells with
the age of each initial cell in the interval $[1,50]$ drawn
randomly from a uniform distribution. Taking into account the
volume of a single cell and the volume of related intercellular
space, this corresponds to a class I spherical volume of
approximately $500 \mbox{mm}^{3}$. The cell division algorithm was
designed with high computational efficiency. To optimize the
algorithm and make modifications to allow for more memory
efficient data storage, the colony is arranged as an array for an
efficient way to store, access and manipulate data as the tumour
cells die, split, mutate and accumulate in the ensemble. However,
the method does not allow us to work with geospatial clustering
effects due to its lack of dimensionality but allows for rapid
computation.

The hadrontherapy begins with a virtual tumour colony with a
pre-defined death probability and hypoxic status. Cells continue
to divide between the fractions during the simulated hydrontherapy
and fractionated treatment continues until all tumour cells have
been eradicated or the desired number of treatment fractions has
been delivered. Cell death is accessed for each cell in the array
for each dose fraction. After cell death, links in the array are
updated to maintain the consecutive order of the cells in the
relevant hourly list as well as the number of living cells. The
emptied array elements are auto-reused after cell death for
efficient computer memory usage. We observed a computation time of
2-3 minutes/trajectory to run an ensemble of $10^{6}$ cells with
fractionated hadrontherapy, with the exception of rapidly
repopulating tumours.

\subsection{Input Parameters and Simulation Details}

The hybrid algorithm contains a number of input parameters which
intuitively describe physical and biological effects of the
interaction of light ions with tumour cells. These parameters
include natural death rates, natural cell repair probabilities,
cell division and multiplication probabilities, spontaneous
mutation rates, etc. The exact values of these parameters can be
taken from experimental results, if identified, otherwise the
parameters are simulated. However, the experimental determination
of some of these parameters is rather challenging and in many
cases, it involves the assumption of a model relating a measured
quantity to the quantity to be determined. The input parameters
were applied in the model after analyzing for physical observables
over a reasonable range of estimated values or probability
distributions. Since the key input parameter values often vary
during treatment simulations, a graphical user interface (GUI) was
developed within the Python environment to enable convenient
values of the desired input parameters. This also allows the
running of multiple batches run iterating over different random
seed numbers and ranges of input parameter values.

To generate a random integer sequence, a set of random floating
point number between $0$ and $1$ with a uniform, normal, lognormal
or exponential distribution, we utilized the Ziggurat pseudo
random number generator \cite{Mars2000}. The object-oriented
programming language Python was used to create the model
algorithm. The numerical loops contain the trees of probabilities
drawn from several probability distributions describing various
biological effects. Following \cite{Fornalski2014} we calculate,
for time step $\Delta t$, the probability that each cancer cell
(i) is irradiated by a hadron or ion; (ii) is killed by a single
precise hit of the radiation; (iii) naturally dies; (iv) naturally
multiplies - the daughter cell is also cancerous; (v) dies due to
cell's radiosensitivity; or (vii) remains unchanged. In case the
cancer cell is not irradiated, it can die naturally, multiply or
stay intact. For each Monte Carlo run, the tumour cell system
evolves in time according to the probabilities defined above. The
Pseudo random numbers are generated to sample probability
distributions based on model calculations describing the above
events.

The trees of probabilities of the above seven scenarios form input
data and are presenting in Tab. \ref{tab1}. Some of the
probabilities, e.g., the probability of natural death of a cancer
cell, probabilities of the division of cancer and mutated cells,
can be approximated by constant values extracted from the
experimental data or simulation results. Remaining probabilities
in the above net describing the biological and physical effects of
the interaction of ionizing radiation with cancer cells are drawn
from probability distributions. For example, The distribution of
hits is given by the binomial distribution, which for a large
number of cells may be approximated by the Poisson distribution.

The probability of hitting a cancer cell by radiation can be drawn
from the binomial distribution, which for a large number of cells
may be approximated by the Poisson distribution, as well as from a
stretched exponential distribution. However, we draw the
probability , $P_{hit}$, form a quasi-linear probability
distribution, $P(D)=1-e^{-c_{1}. D}$, where $D$ is the dose per
single cell per time step and $c_{1}$ is a scaling constant. Such
distribution can also be used for the death of irradiated cancer
cells because of its specific radiosensitivity, $P_{CRD}$. On, the
other hand, the probability distribution for cancer
transformation, $P_{RC}$, is well described by sigmoidal
probability distribution \cite{Cohen2011} with critical index $n$,
$P(Q)= 1-e^{-aQ^{n}}$, where $Q$ is the number of mutations and
constant $a$. This distribution, however, does not take into
account the cancer-cell microorganism mutations that arise
non-homogeneously over time.

The tumour was exposed to hadron and ion radiation dose of 2 - 5
Gy per day up to a maximum dose of 10 Gy. By varying the random
seed number in the hadrontherapy algorithm, we generate six
ensembles of cell irradiation measurements for each dose value for
later analysis. The expectation values and statistical error
estimates of the observables from Monte Carlo simulations were
estimated using the jackknife method. The statistical errors were
estimated by grouping the stored measurements into 5 blocks, and
then the mean and standard deviation of the final quantities were
estimated by averaging over the "block averages", treated as
independent measurements. The estimate of the error of observable
$\rho$ was calculating by using
\begin{displaymath}
\delta \rho = \sqrt{\frac{M-1}{M}\sum_{m=1}^{M}(\bar{\rho}_{m}
-\langle \rho \rangle )^{2}}.
\end{displaymath}
Statistical significance between two data sets was accessed using
a 2-tail $t$-test with a $95\%$ confidence interval.

\begin{table}{!ht}
\begin{center}
\caption{Input parameter values tumour growth, radiation-induced
DSBs for V79 and probabilities used for the simulations.}
\label{tab1}
\begin{tabular}{ccccccc}\hline\hline
Parameters (Ref) & Value \\ \hline
{\emph{Tumour growth}} & \\
Number of cancer cells & $10^{6}$ \\
Cell diameter & 20 $\mu$m \cite{Harting2010}\\
Radius of capillary cells & 10 $mu$m \cite{Harting2010}\\
$Po_{2}$ threshold for hypoxia & 5 mmHg \cite{Harting2010}\\
Maximum $OER$ ($p$) & 3 \cite{Hall2000}\\
Maximum $K$ at $OER$ & 3 \cite{Hall2000}\\
{\emph{Radiation-induced DSBs for V79 cell}} \cite{Wang2018}& \\
$\mu_{x}$ & 0.956(23) \\
$\mu_{y}$ & 0.030(17)\\
$\zeta$ & 0.041(20) \\
$\xi$ & 0.060(38) \\
$\eta\lambda_{p}\rightarrow 1$ & $9.78(10)\times 10^{-4}$\\
$\eta\lambda_{p}\rightarrow \infty$ & 0.0065(1)\\
{\emph{Probability distributions}}\cite{Fornalski2011,Fornalski2014} & \\
Spontaneous mutation in a cell & $(1-\tau )(1-e^{-aK^{n}})+\tau$ \\
Natural repair of one mutation & $\delta e^{-aK^{n}}$ \\
Mutation develops in the irradiated cell & $1-e^{-const \cdot D}$ \\
Mutated cell $\rightarrow$ cancer cell & $1- e^{-const \cdot D}$
\\\hline
\end{tabular}
\end{center}
\end{table}

\section{Results and Discussion}

\subsection{Yield of radiation-induced DSB and $\alpha /\beta$ ratio}

The DNA damage induced by ionizing radiation was obtained in terms
of isolated and multiple strand breaks. Fig. \ref{fig1} shows the
dependence of the yield of radiation-induced DSB
($\mbox{Gy}^{-1}\mbox{cell}^{-1}$) of V79 cells on LET. We notice
that at low and medium LET, DSB yields generally increase with
increasing LET, from $\sim 45$ per Gy per cell at $1.2$
keV/$\mu$m, up to $98 - 102$ per Gy per cell at $300 - 550$
keV/$\mu$m. The calculated yield of DSBs after irradiation with
protons showed an interesting feature with LET around $15$
keV/$\mu$m. At LET below $12$ keV/$\mu$m, the DSB yield per track,
$\lambda$ increases sharply with LET, however, $\lambda_{p}$, the
average number of DSBs yielded by each primary particle that
causes DSB increases rather slowly and is nearly similar to that
with $\gamma$. This may explain the reason behind nearly similar
biological effectiveness of low LET protons and photons. For
proton LET above $12$ the total yield showed a decrease of about
$15\%$ with increasing LET within the range $13 - 52$ keV/$\mu$m.
Such effects were also observed in other studies with varying
percentages of yield reduction \cite{Nikjoo2001,Friedland2003}.

\begin{figure}[!ht]
\begin{center}
\scalebox{0.60}{\includegraphics{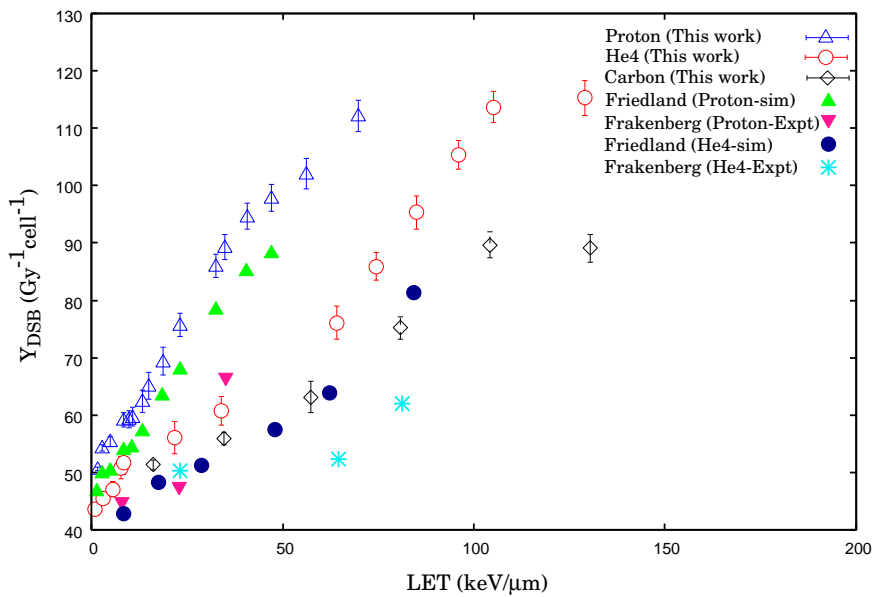}}
\caption{\label{fig1} Yields of radiation-induced DSB with protons
(open triangles), He4 (open circles) and carbon ions (open
diamonds) as a function of LET. For comparison, we also show the
simulation results obtained by Friedland {\emph{et al}}.
\cite{Friedland2017} using PARTRAC and experimental results of
Frankenberg {\emph{et al}}. \cite{Frankenberg1999}}
\end{center}
\end{figure}

At high LET values, the total yield of breaks shows the
ion-specific behaviour; for instance, at 60 keV/$\mu$m, protons
are fairly dominant in inducing DSB ($\sim 102
\mbox{Gy}^{-1}\mbox{cell}^{-1}$) compared to helium nuclei ($\sim
87 \mbox{Gy}^{-1}\mbox{cell}^{-1}$) and carbon ions ($\sim 75
\mbox{Gy}^{-1}\mbox{cell}^{-1}$). This reduction can be attributed
to the decrease in the indirect contribution with increasing LET,
which is likely due to the structure of the proton track.

A comparison between the yield results obtained in this work and
other simulations and experimental data found in the literature
was performed. Our results for proton and $\alpha$-particles DSB
yields follow the same behaviour (DSB yield increasing with LET,
for both proton and $\alpha$-particles) as in other simulations
\cite{Wang2018,Meylan2017}. Such behaviour may be related to the
increase of the clustering of energy depositions and chemical
species production. Our results for protons are consistent with
those reported by Friedland {\emph{et al}}. \cite{Friedland2017}
where PARTRAC has been used to simulate track structures of
protons, $\alpha$-particles and light ions with low to medium
energies. In terms of absolute values, there are discrepancies of
less than $2\%$ between these two works for DSB yield due to
protons indicating that MCDS algorithm gives reliable results of
the damage yields that are comparable to those obtained from
computationally expensive but more detailed track structure
simulation models. This is as expected since the MCDS simulations
implicitly account for both direct and indirect DNA damage
mechanisms. For clarity of data points in the effective plot, we
do not show the results from these studies. A comparison with the
experimental data, however, shows large discrepancies at both low
and medium LET. This might largely be, among other factors, due to
the dependence of experiments on the ability to determine small
DNA fragments, similar to the influence of physical and biological
models implemented in Monte Carlo algorithms on the simulations.

With the values for DSB yield established for specific hadron
spices, we obtain the data for $\alpha$ and $\beta$ parameters of
the model and show their dependence on LET in Fig. \ref{fig2}. The
surface plots (panels a-d) of linear-quadratic parameters with
protons and helium ions in Fig. \ref{fig2} show different trends
with increasing LET and dose. It can be seen that the
linear-parameter, $\alpha$, becomes progressively steeper as the
particle LET increases. At a given LET, the alpha parameter for
protons (Fig. \ref{fig2}e) is higher than that for the helium and
carbon ions. This could be due to increases in proton relative
effectiveness for DSB induction in the LET range 5-70 keV/$\mu$m
caused by a difference in the track structures of the ion beams at
the same LET through differences in the effective charge and the
velocity of the ion. With helium and carbon ions, the linear
parameter seemed to increase to a maximum before starting to fall
at high LETs. The position of the maximum alpha shifts to higher
LET values for carbon ions. The modelled alpha values show a good
agreement with the LEM-based results \cite{Giovannini2016},
particularly up to approximately 12 keV/$\mu$m using protons.
Beyond this value, the LEM exhibits an enhancement in the linear
parameter. Nonetheless, while our parameters $\alpha$ results show
a slow increase, compared to LEM model, both show a similar trend
as expected. As a result, our $\alpha$ results with proton
irradiation are more reliable for low LET values than for high
ones. No relevant differences for dose per step and aerobic and
hypoxic conditions are apparent up to 8 keV/$\mu$m. The $\alpha$
values for helium and carbon-ions show a similar trend with nearly
the same values at similar LETs under both aerobic and hypoxic
conditions.

\begin{figure}[!ht]
\begin{center}
\begin{tabular}{ccccc}
\scalebox{0.45}{\includegraphics{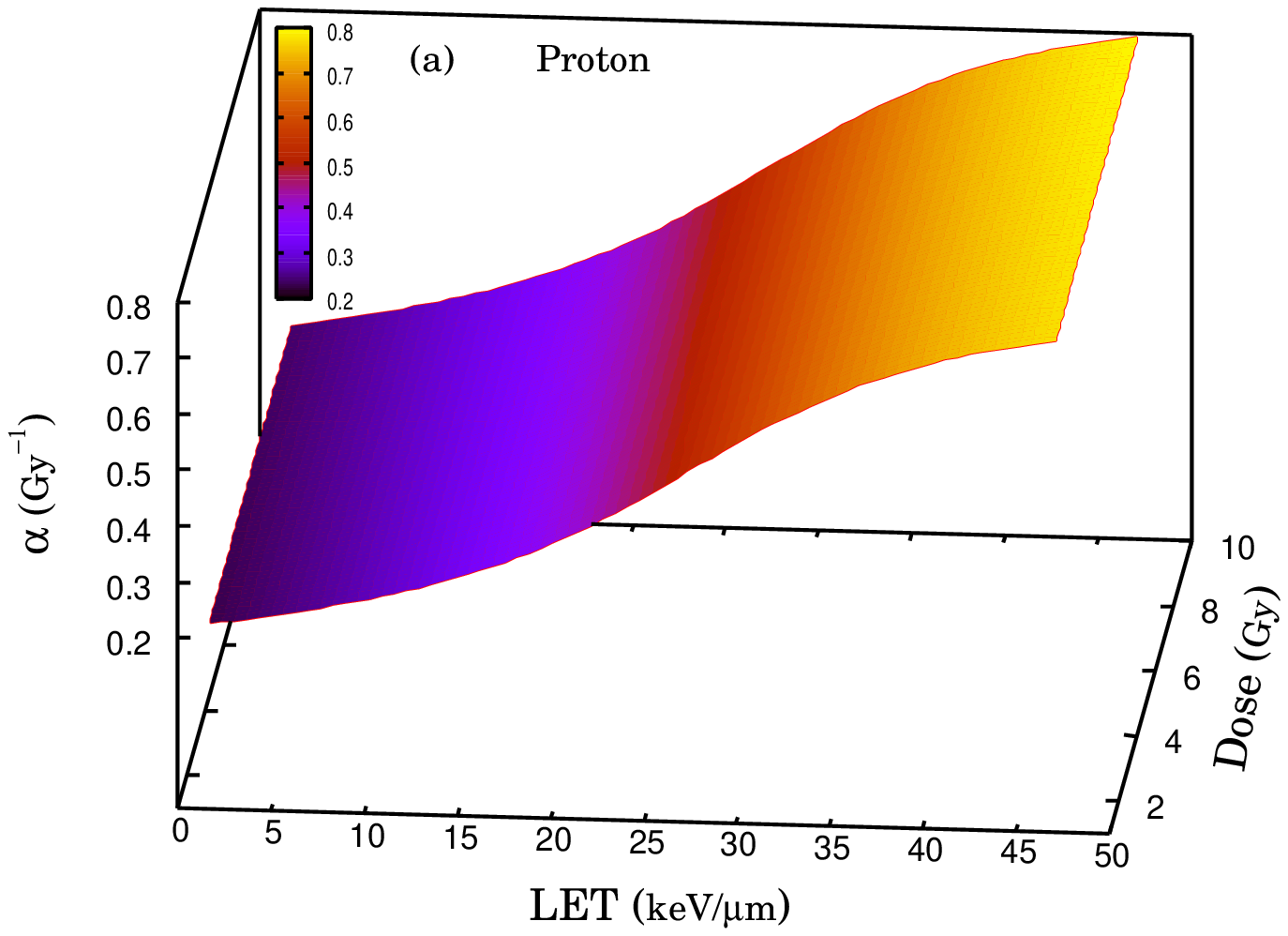}} &
\scalebox{0.45}{\includegraphics{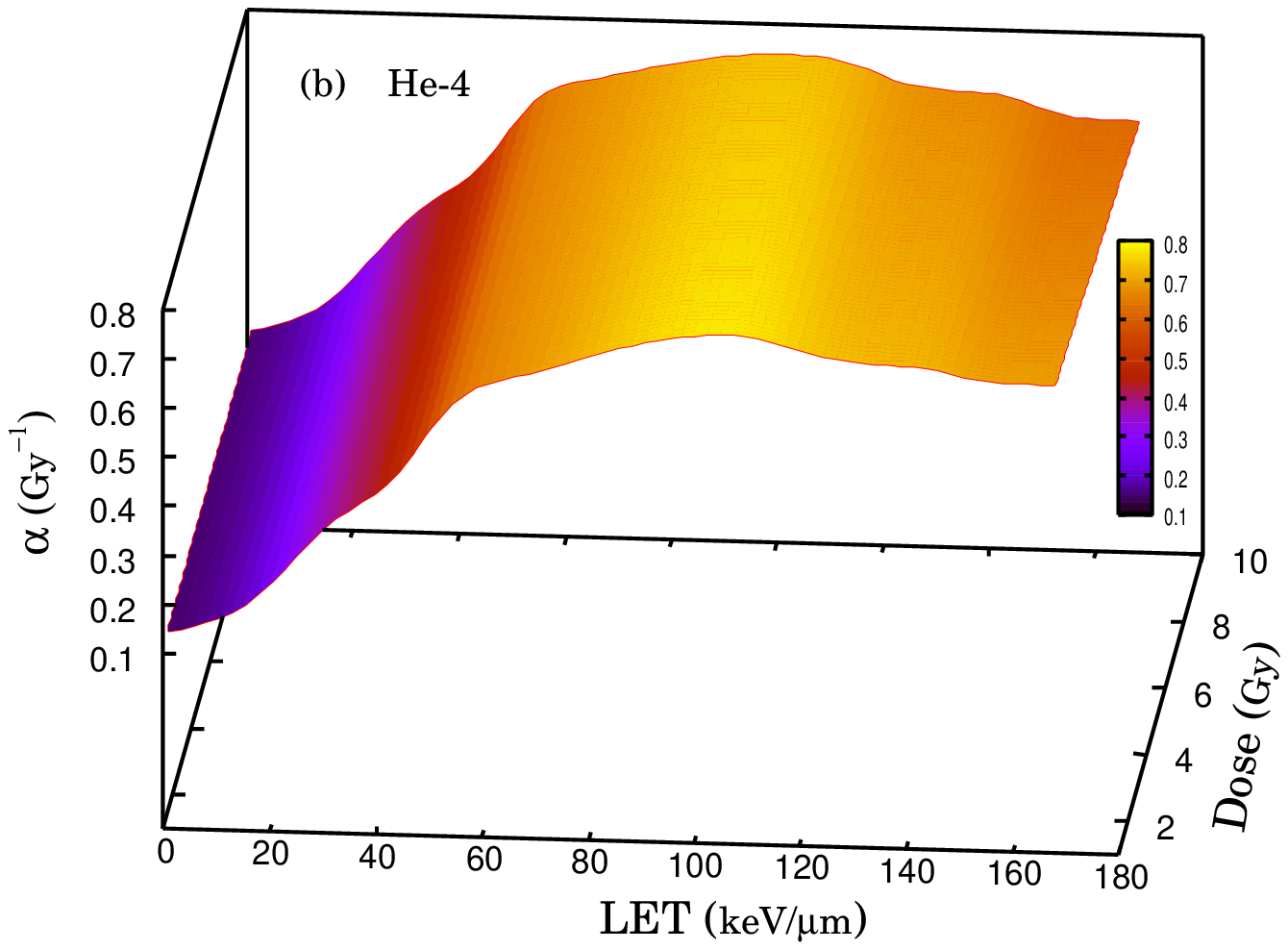}}\\
\scalebox{0.45}{\includegraphics{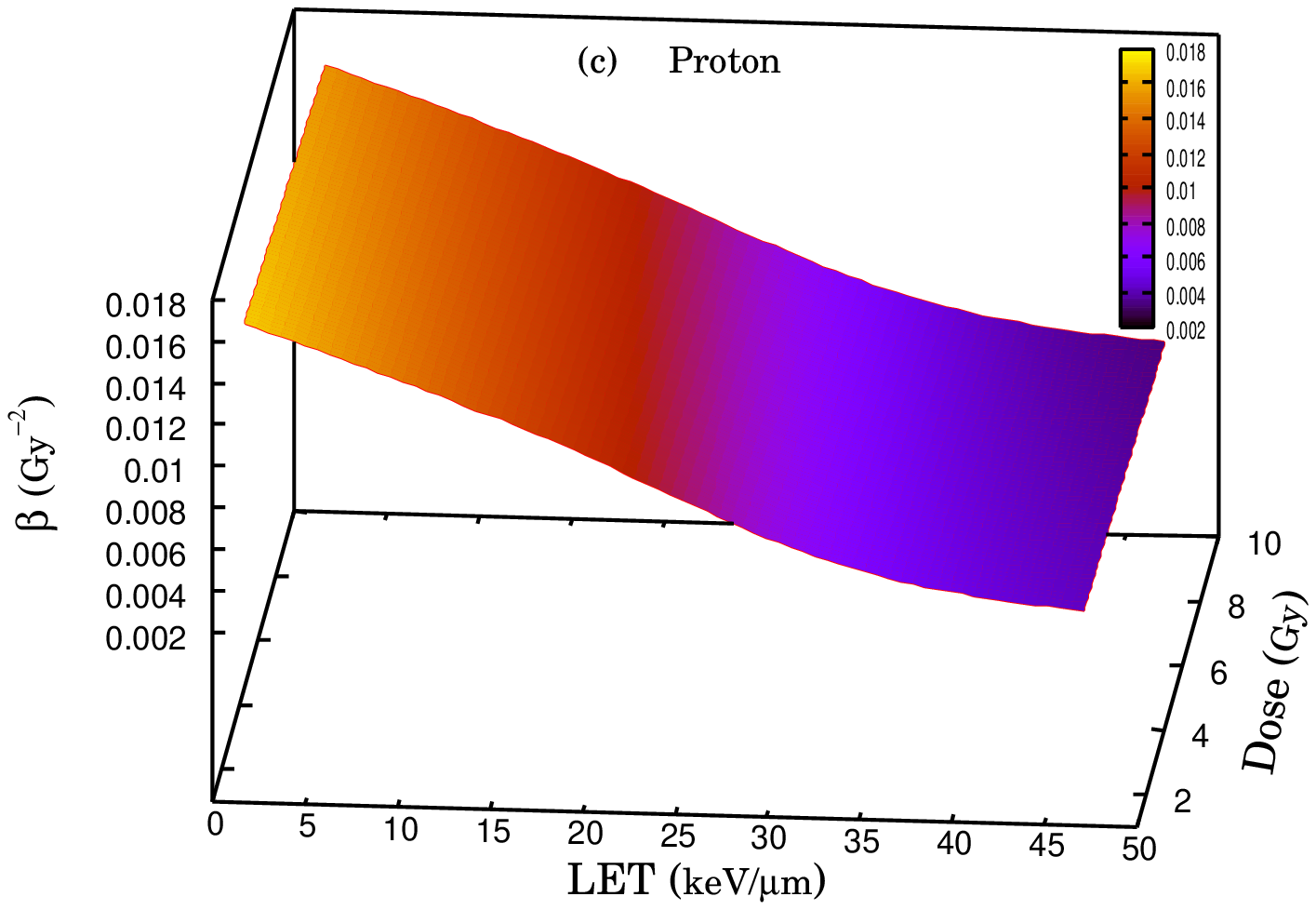}}&
\scalebox{0.45}{\includegraphics{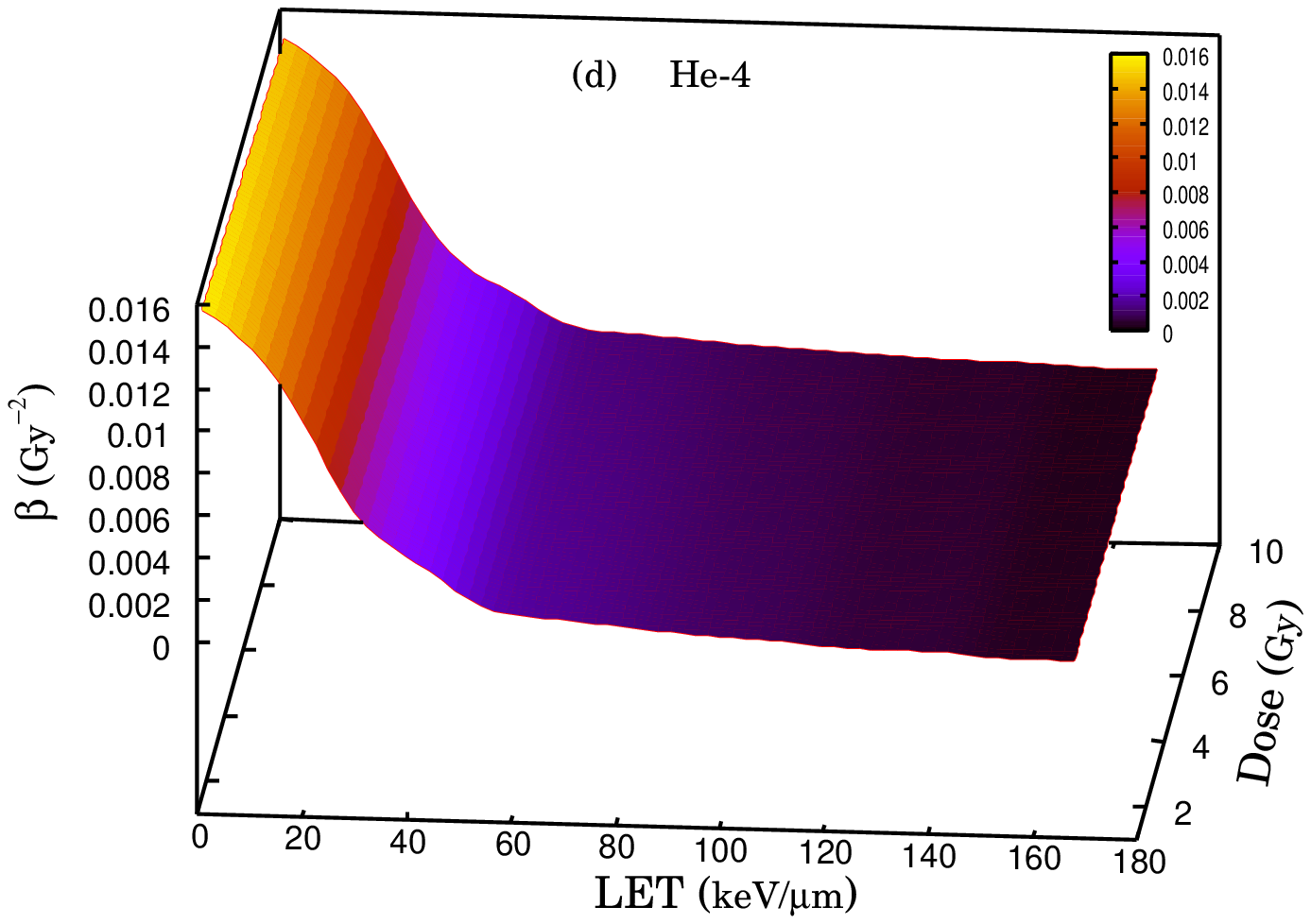}}\\
\scalebox{0.32}{\includegraphics{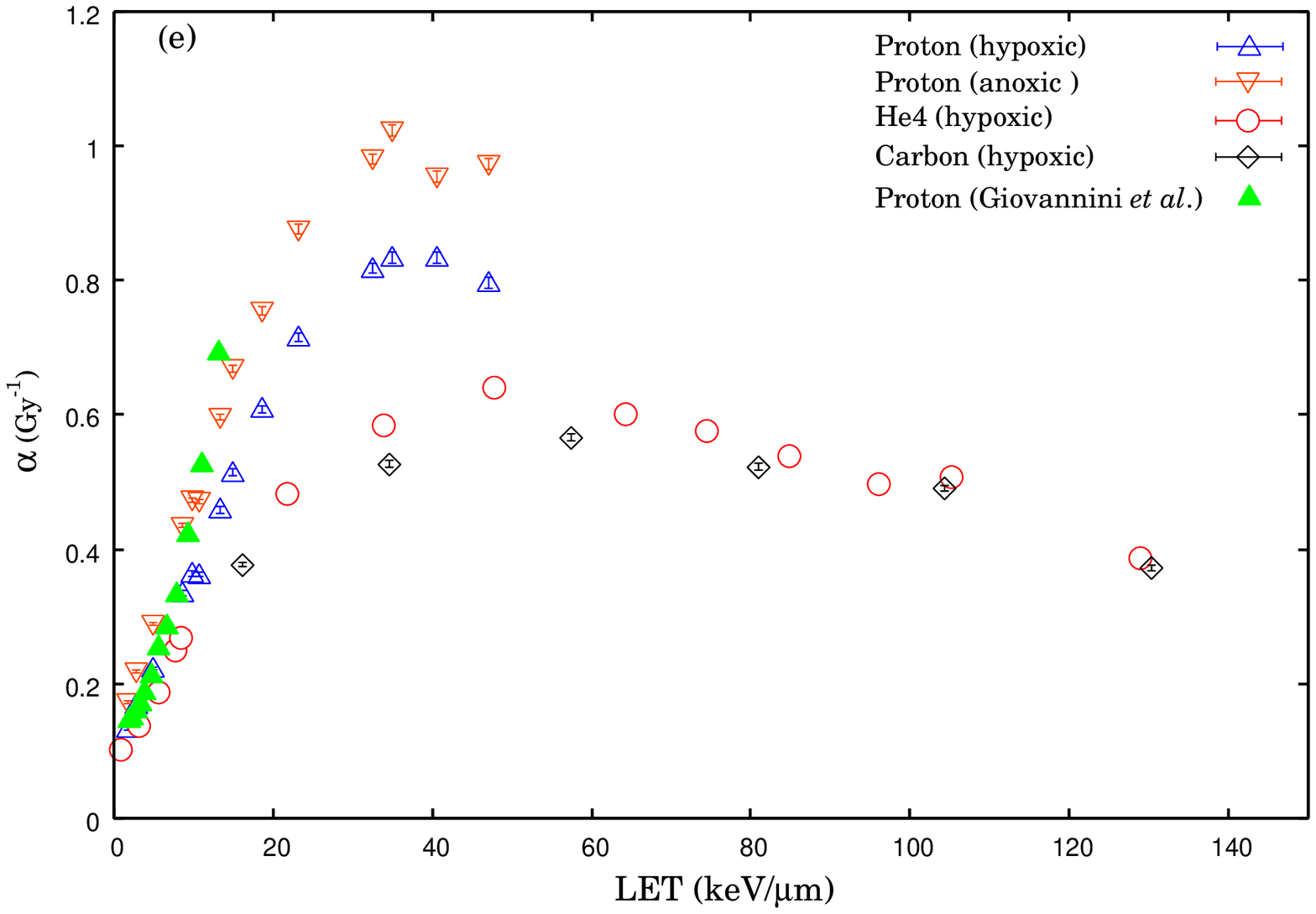}}&
\scalebox{0.32}{\includegraphics{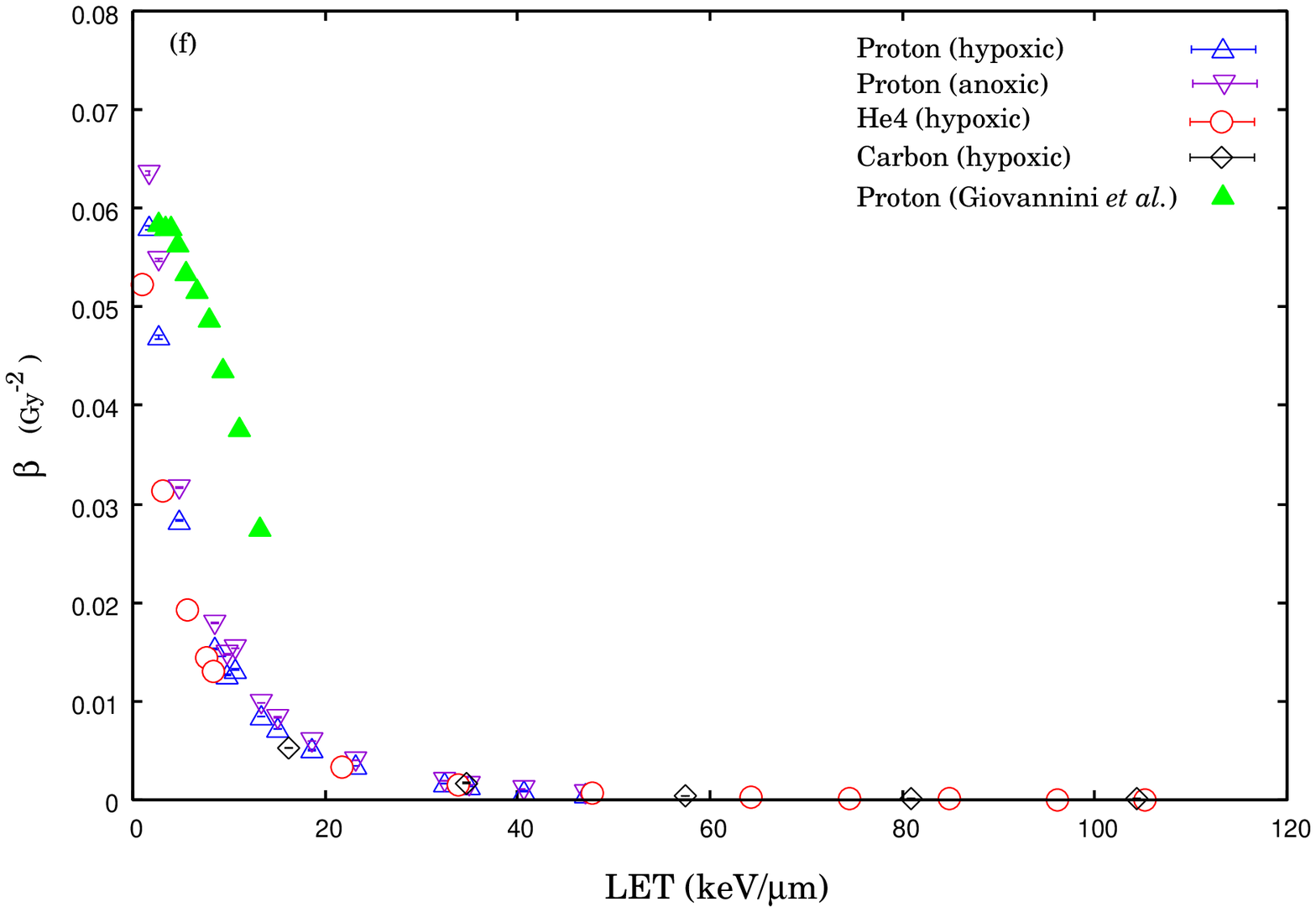}}
\end{tabular}
\caption{\label{fig2} Linear-quadratic parameters of the model for
exposed to protons (panels a and c) and He4 ions (panels b and d)
for a range of LET values. Comparison with the results obtained
using LEM model \cite{Giovannini2016} is shown in panels (e) and
(f). }
\end{center}
\end{figure}

For medium and high LET values, our simulation results show an
almost vanishing $\beta$ parameter for all radiation types used in
this study (Fig. \ref{fig2}f). For helium and carbon-ions, we
observe the general trend, an increase at low LET values followed
by a decrease at higher LETs, showing that the trends are
qualitatively similar across cells of different
radiosensitivities. Similar to the $\alpha$ maximum, the fall-off
for $\beta$ is shifted to higher LET for carbon-ion. Again, in
terms of the absolute values, the variation trend of our estimated
with those of LEM model is significant at low LET values under
both aerobic and hypoxic conditions. This may be due to an
improved description of the quadratic parameter within the LEM
framework \cite{Friedrich2012} as well as in this study
(Eq.\ref{eqn3}), where the variations of dose affect $\beta$ more
than $\alpha$, especially at high LET values.

Since the $\alpha /\beta$ is mainly attributed to the number of
primary particles that cause DSBs per dose for charged particles
at higher LET values, the cell sensitivity has less effect of cell
killing for charged particle species. Fig. \ref{fig3}a shows the
ratio (on the logarithmic scale) for cells irradiated by various
hadrons at different LETs. The initial slope of the ratio increase
steeply with LET. This increase is due to an increase in $\alpha$
value (primarily due to the clustered DNA damage effect) with LET
as well as the decrease in the interaction of DSBs (contribution
to $\beta$ term of the mechanistic model) induced by different
primary particles. This interaction becomes vanishingly small at
intermediate LET values. The ratio seems to reach a plateau for
high LET values with helium- and carbon ions. This is more likely
due to the saturation in the clustered DNA damage and the effect
of overkill on cell death.

The question of whether the model parameter are interdependent is
an important indicator that needs to be explored. To explore the
correlation between the model parameters, we plot the quadratic
parameter against linear parameter for different radiation types
at different LET-dose combination in Fig. \ref{fig3}b. Whereas, a
plateau with no visible interdependence, between the parameter
with carbon-ions is observed, the plot suggests a clear negative
correlation between the parameters with protons and helium-ions.

\begin{figure}[!ht]
\begin{center}
\begin{tabular}{ccccc}
\scalebox{0.33}{\includegraphics{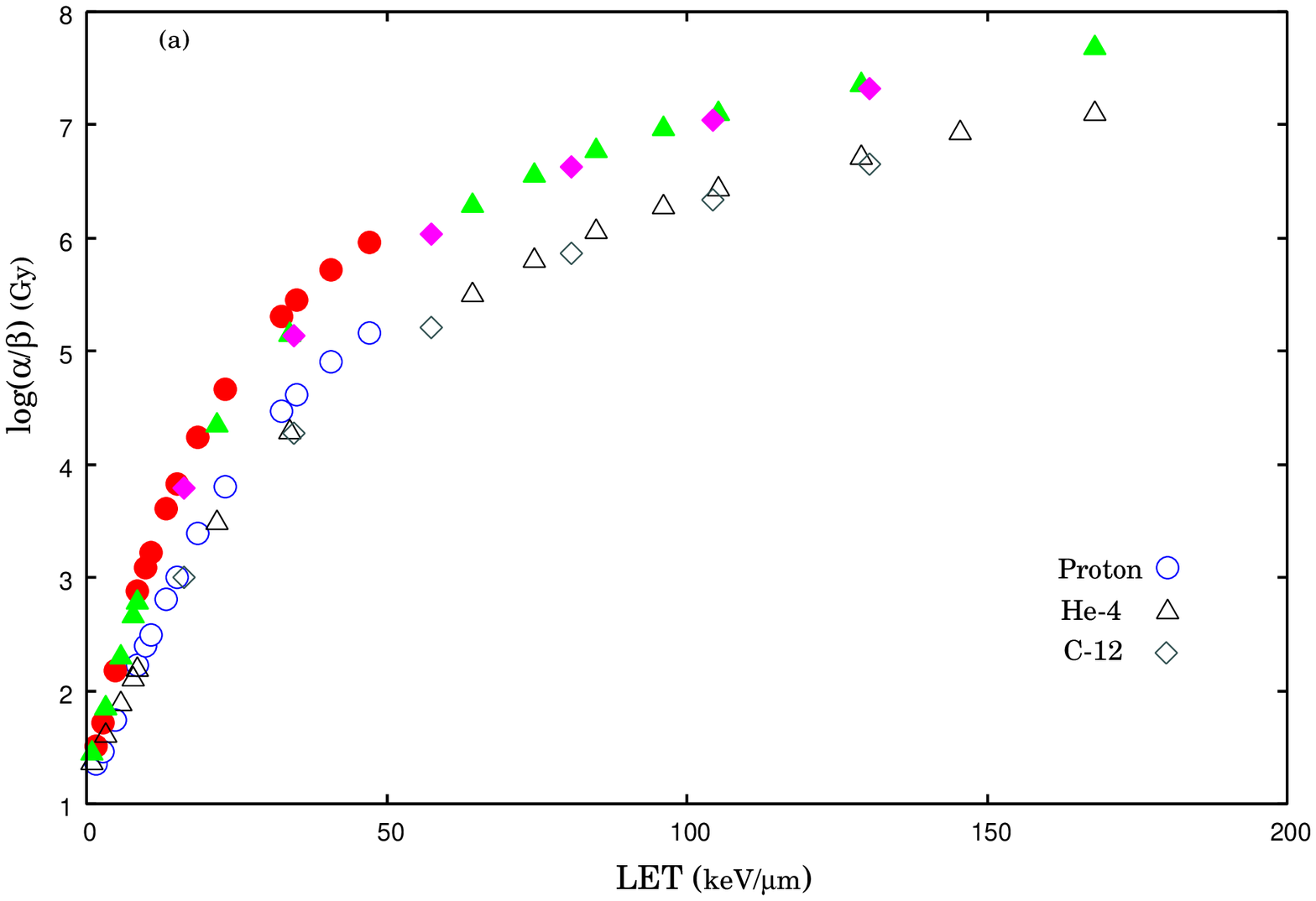}} &
\scalebox{0.33}{\includegraphics{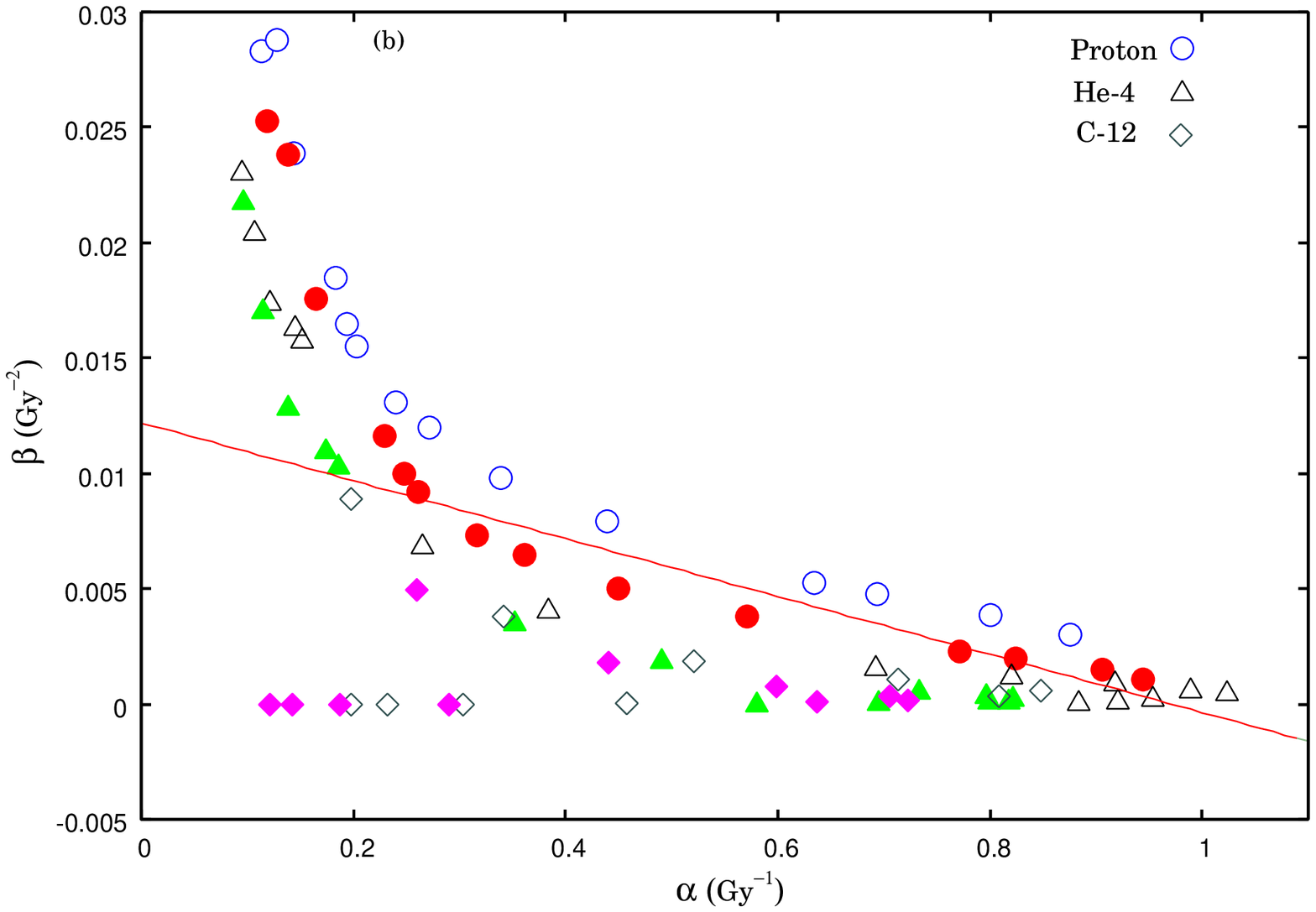}}
\end{tabular}
\caption{ \label{fig3} (a) $\alpha /\beta$ (in logarithmic scale)
as a function of LET. (b) Correlation between LQ parameters of the
cancer cells. The open symbols represent the data using protons
(open circles), helium-ions (open triangles) and carbon-ions (open
diamonds) at 3 Gy dose. The corresponding solid symbols represent
the data at 5 Gy. The straight line represents the linear fit to
the data using proton irradiation (solid circles).}
\end{center}
\end{figure}

For protons and helium, the signal is noisy at earlier $\alpha$
values, and hence we fit a linear regression to the data in the
interval $0.3 \leq \alpha \leq 1.0$. The best fit curve to the
data gives a negative slope of $(-0.0125\pm 0.0013)$ and has
$\chi^2 /N_{df} = 0.82$. The fit to the data using helium-ions
gives a smaller slope of $(-0.0060 \pm 0.0014$. The statistical
uncertainties on our measurements are typically on the few percent
levels. In both the cases, a $p$-value test, at $5\%$ significance
level, was performed on the correlation coefficient between the
model parameters with $r_{p}= -0.8679$ ($p=0.000057$) and $r_{He}=
-0.9162$ ($p=0.000002$), respectively. These values suggest a
correlation between the linear and quadratic parameters for proton
and helium-ion irradiation. We will keep a close eye on the trend
of the survival curves to see if such an interdependence
influences the steepness in the exponential fall of the curves.

\subsection{LET-dependent RBE}

Whereas a large amount of data from different experimental
protocols and biological models are available
\cite{Friedrich2012b}, the adoption of a simple and unique RBE-LET
relationship in effective treatment planning is surrounded by a
number of uncertainties. Few studies have supported a reasonable
approximation of fixed RBE to describe the increased effectiveness
of light ions
\cite{Paganetti2002,Paganetti2006,Giovannini2016,Leeuwen2018}, the
concerns for a better understanding of RBE-LET relationship for
significant clinical relationship have been raised. To further
explore the impact of LET on radio-sensitivity, we analyse and
display in Fig. \ref{fig4}, the RBE corresponding to the initial
slope for different particle types, with x-rays as reference
radiation ($alpha_{x}= 0.616 GY^{-1} \beta_{x}=0.062 GY^{-2}$). It
is obvious from the effective plots (a-c) that the maximum in RBE
depends on the particle species, where heavier particles have the
maximum at higher LET values and that the lighter ions provide
higher RBE values for a fixed LET. It can be seen from Fig.
\ref{fig4}c that RBE decreases with increasing dose. Comparison
with other numerical estimates and experimental results shows a
good agreement, especially for protons. However, we notice a
deviation of our results at high LET value for helium ions, but
the overall trend agrees well with those of the observed results
for different particle species at different LET.

\begin{figure}[!ht]
\begin{center}
\begin{tabular}{cccccccc}
\scalebox{0.33}{\includegraphics{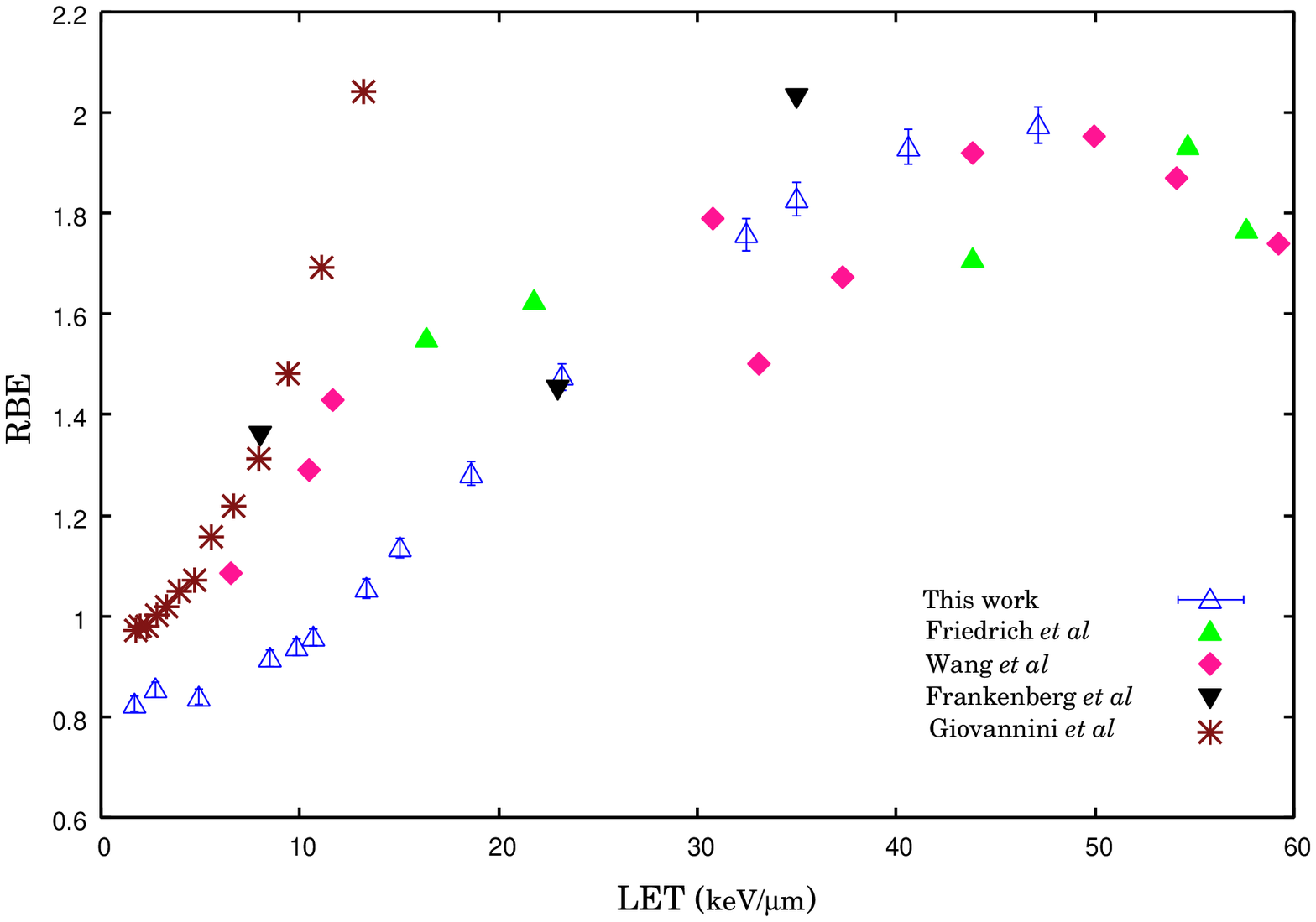}} &
\scalebox{0.33}{\includegraphics{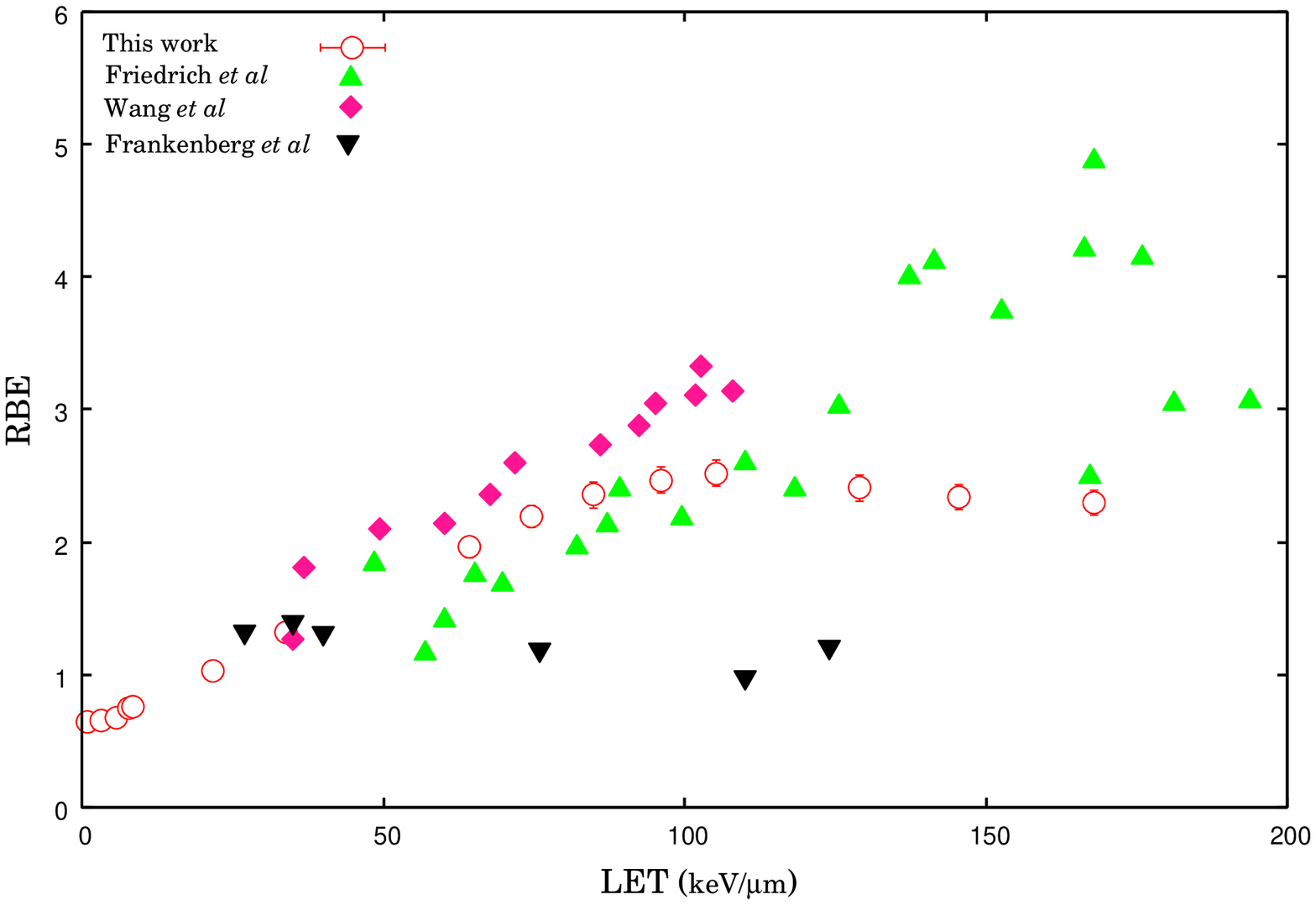}}\\
\scalebox{0.33}{\includegraphics{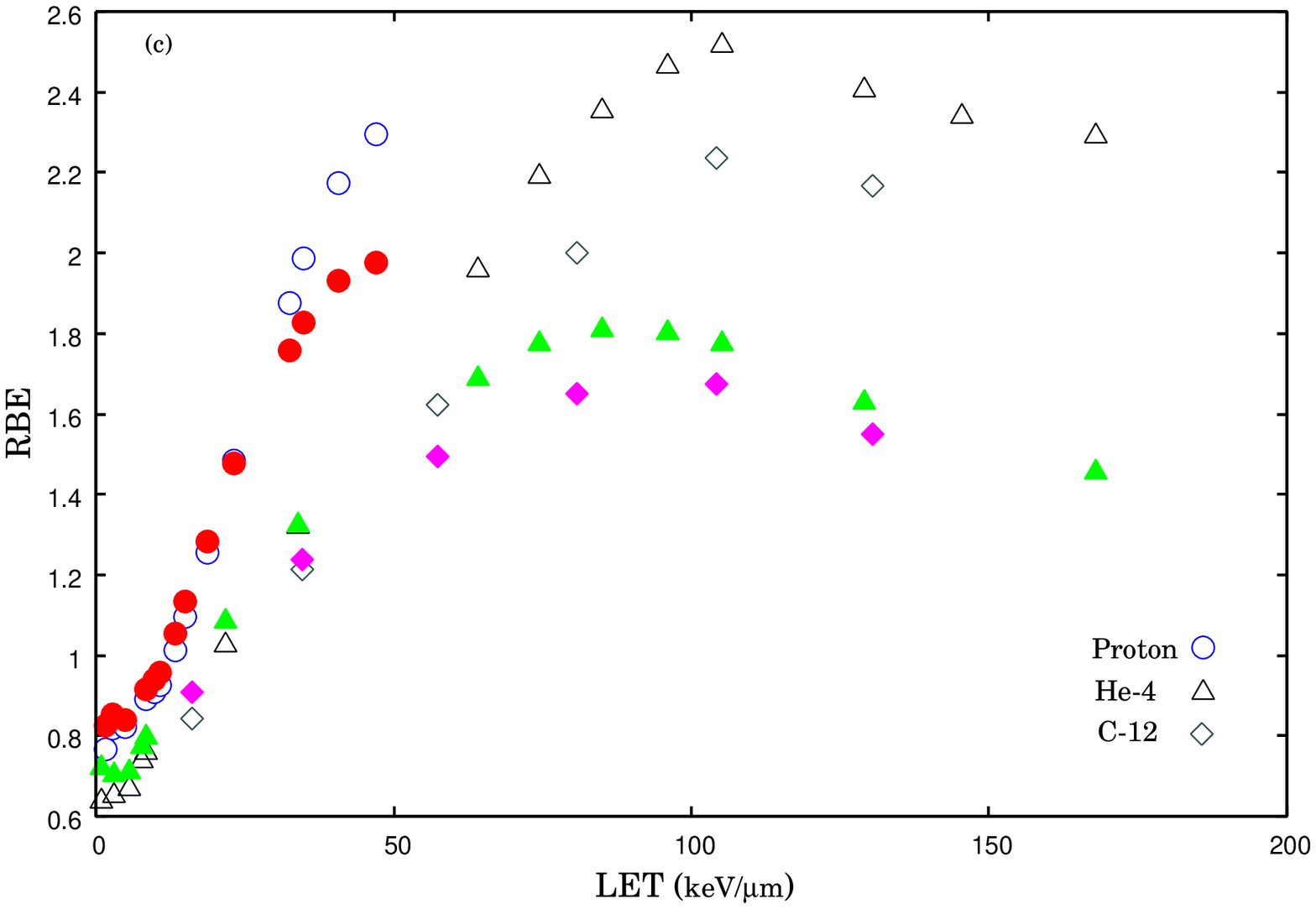}}&
\scalebox{0.33}{\includegraphics{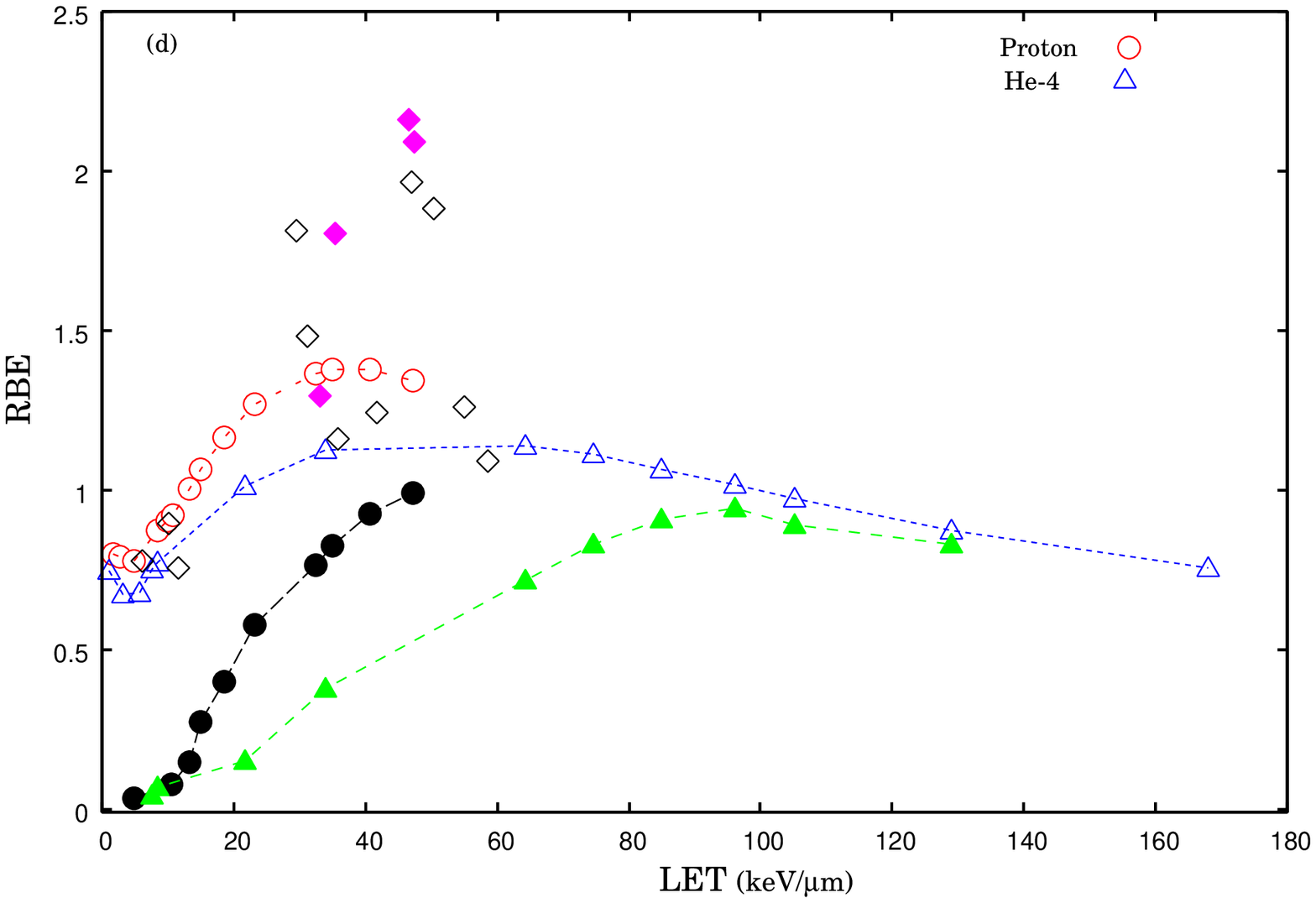}}
\end{tabular}
\caption{\label{fig4} LET-dependent simulated values of RBE of
protons (panel a), helium-ion (panel b). Also, are shown the RBE
values obtained from earlier studies for comparison
(\cite{Frankenberg1999,Friedrich2012b,Wang2018}. Panel (c) shows
the RBE values at different doses (open symbols - 3 Gy and filled
symbols - 5 Gy). Panel (d) shows RBE values of ionizing radiations
at $10\%$ survival level in comparison with values obtained in
full survival level.}
\end{center}
\end{figure}

To explore how RBE compares in the limit of full survival level
and $10\%$ survival dose, we collect and display RBE values
obtained from the $10\%$ survival when cells are irradiated by
different particle species with different LET in Fig. \ref{fig4}d.
We notice that the RBE-LET spectra are different for different
particle types; RBE with protons increases with LET, peaks at
around 45 keV/$\mu$m and then decreases with LET, whereas, for
helium-ions, the RBE increases slowly in the medium LET region
with a rather broad peak in the near-high LET region. The RBE
values show a good agreement with the measured values at small LET
for protons but considerable scattered of the experimental values
around our numerical estimates with small and medium LET values
for helium-ions. The RBE for $10\%$ survival level plots with
protons and helium-ions nearly level up in the high-LET region and
approach approximately to 1.

As far as the role of sensitizers, such as oxygen, on the
radioresistance of the cells is concerned, we extract $\alpha$ and
$\beta$ from the slopes of the survival curve at $10\%$ survival
level to calculate OER values using Eq. (\ref{eqn5}). We notice
that a decrease in the OER values for helium and carbon-ions
starts at around 50 keV/mm, passing below 2 at around 100 keV/mm,
and then reaches approximately 1 (significantly lower for
helium-ions) in the very high-LET region. The OER was
significantly lower for helium ions than the others. The presence
or absence of oxygen mainly affects the initial radiation-induced
DSB yield and not the rate of DSB rejoining. However, due to the
dominant contribution of direct effect in the high LTE region and
available oxygen-independent pathway, the dependence on oxygen for
cell kill becomes less important.

We conclude this section by exploring the dependence of RBE on
cellular repair factor and the radiation type. Note that for
smaller values of $\beta$/$\alpha$ the RBE increases linearly with
an initial slope of ($2.11\pm 0.7$) Gy and decreases sharply for
larger values of $\beta$/$\alpha$ with a slope ($-4.11\pm 0.7$).
The considerable scatter of the data points for various particle
types indicate particle specific behaviour of the initial slope of
RBE.

\begin{figure}[!ht]
\begin{center}
\begin{tabular}{cccccccc}
\scalebox{0.33}{\includegraphics{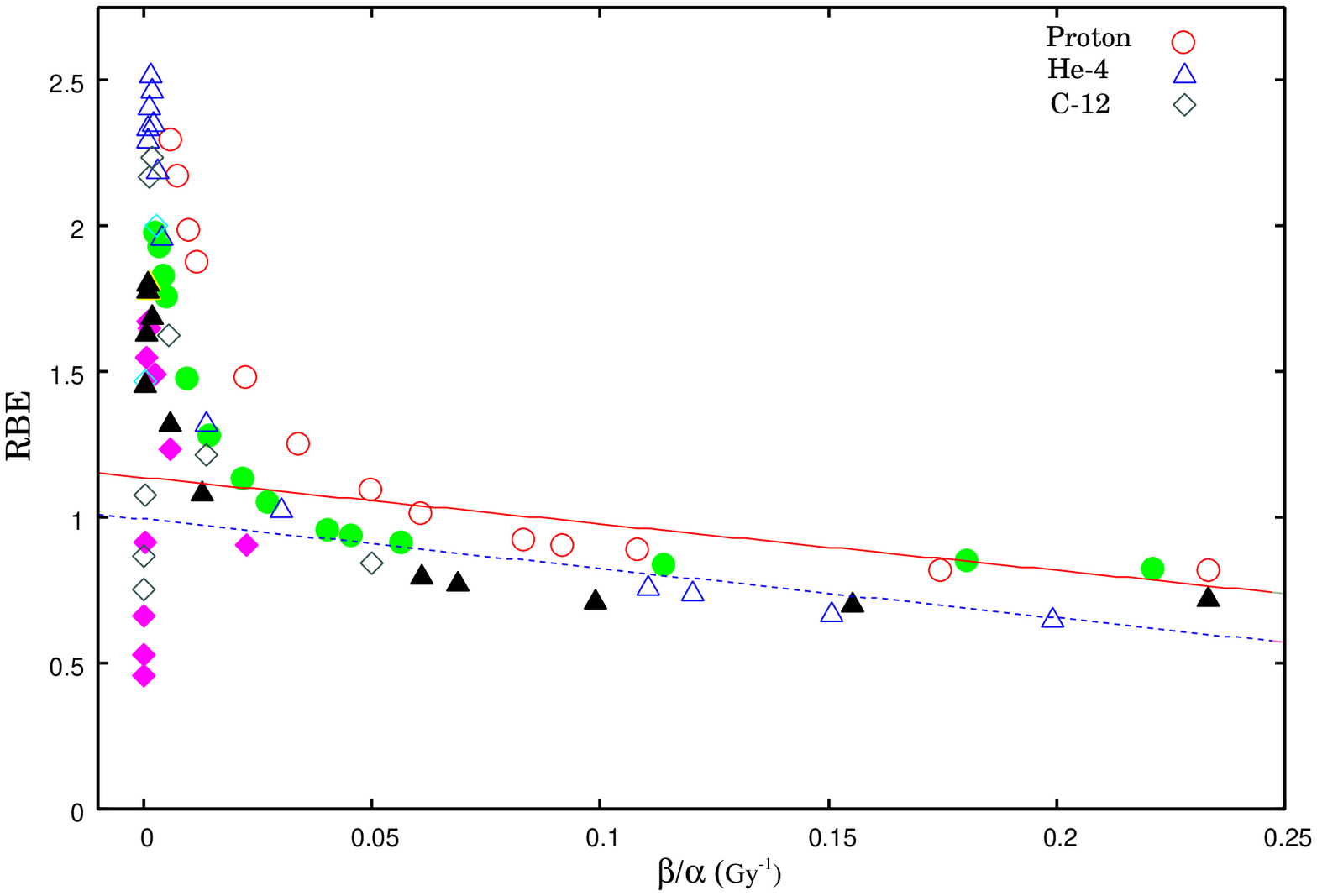}} &
\scalebox{0.33}{\includegraphics{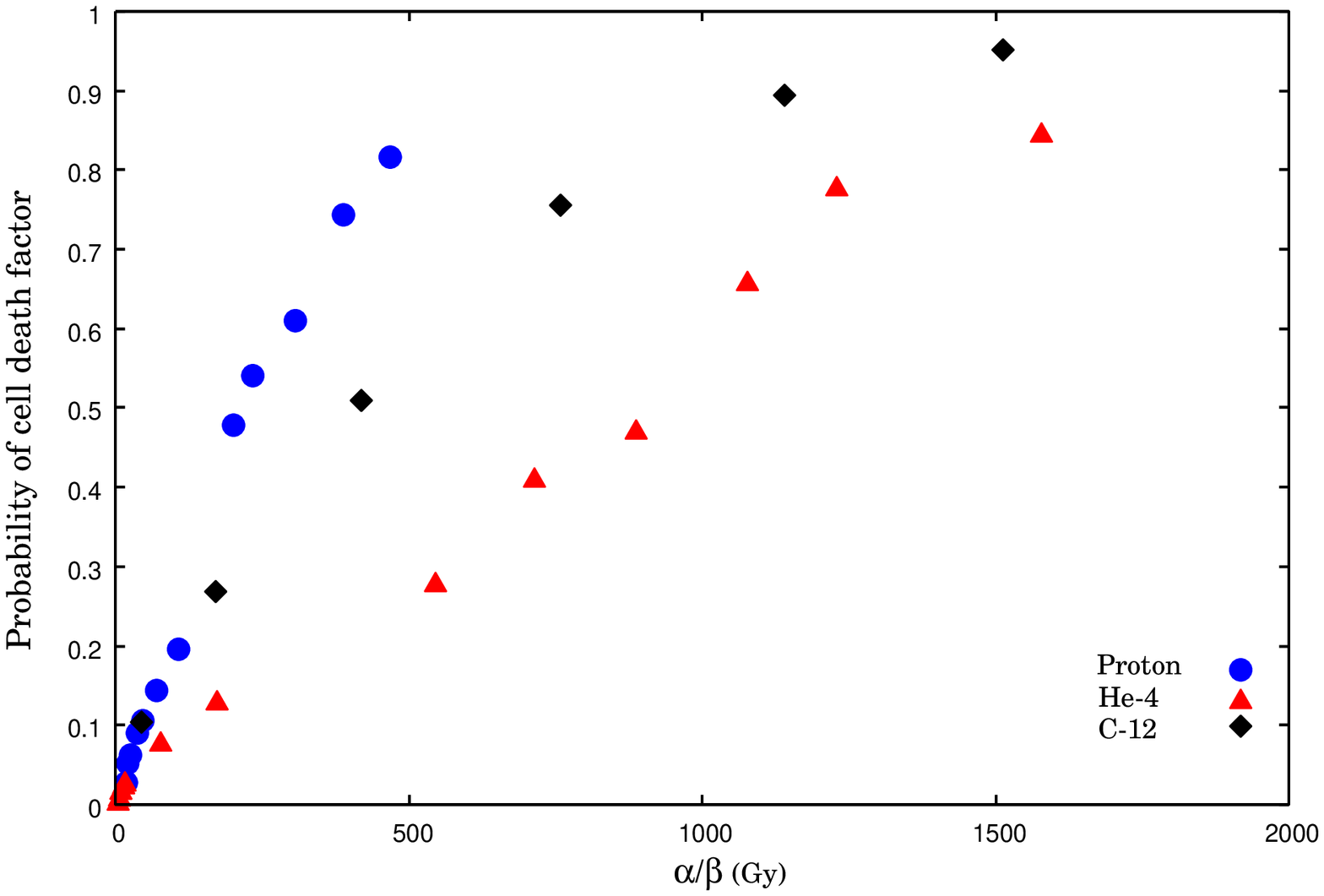}}
\end{tabular}
\caption{\label{fig5} (Left panel) Dependence of RBE values of
ionizing radiations on inverse ratio at 5 Gy dose. The lines are
linear fits to the data using protons (open circles) and He-4 ions
(open triangles). The corresponding solid symbols represent the
data at 10 Gy dose. (Right panel) Probability of cell death factor
with changing dose as a function of cellular repair factor using
protons, helium and carbon ions.}
\end{center}
\end{figure}

The average RBE of protons seems to exceed that of helium- and
carbon-ions at smaller $\beta$/$\alpha$ value for a given LET by a
factor of nearly two. Also, similar to the trend of a shift in
maxima of RBE-LET relationship towards higher particle LET for
helium and carbon-ions (Fig. \ref{fig4}d), the RBE maxima shifts
towards higher values of the ratio $\alpha /\beta$. A linear fit
of the form $RBE_{3Gy}= a+b\cdot (\beta /\alpha)$ gives slopes of
$(-1.59 \pm 0.16)$ and $(-1.69\pm 0.20)$ with
$\chi^{2}/N_{df}=0.74$ and $0.69$ for protons and helium-ions,
respectively. This implies that the cells with higher repair
factor ratio provide large RBEs for medium and larger doses
compared with cells with a smaller $\alpha /\beta$ ratio.
Therefore, tumours with high repair indicators surrounded by
healthy tissue with smaller repair indicators are suited for
hypofractioned regimens than from normal fraction schemes and vice
versa. This clearly indicates that both LET and $\alpha /\beta$
ratio need to be taken as RBE predictors for any hadrontherapy
treatment plan.

Finally, to explore the possibilities in adjusting the
radiosensitivity of cells in the model, we plot, in Fig.
\ref{fig5} (right panel), the probability of death cell factor as
a function of cellular repair factor $\alpha /\beta$. We notice
that, compared to heavy ions, the probability of cell death factor
increases steeply with proton, with differences attributed to
$\alpha /\beta$ ratio, cellular kinetics and to the way the energy
from radiation exposure is deposited.

\subsection{Cell survival curves}

Using pulses of dose for cancer cells, we observed that the cell
colony grew rapidly at low LET ($< 5$ keV/$\mu$m) for dose rate $<
20$ mGy/step for all the three radiation types. This is as
expected since, under the input conditions, the probability
distributions of cell splitting and cell multiplication dominate
and dose pluses are insufficient for killing the cancer cells.
Also, the adaptive response at low LET-dose proved effectively
insufficient to make a contribution to the increase in the
frequency of mutations. Whereas the medium and strong adaptive
response input signals showed a significant change in mutation
frequency, the cell colony continued to grow for low LET-dose
values. Proposed dose and LET-dependent cell survival responses
are simulated by choosing the minimum LET-dose combination for
which the cell survival curves converge after a certain number of
time-steps.

We noticed a clear split of the iteration dose-LET spectra among
protons, helium-and carbon-ions. At LET $\geq 8.5$ keV/$\mu$m and
20 mGy/step dose rate, protons are more effective in cell killing
than helium- and carbon ions. Statistically, all the cancer cells
are killed after the 80th time step with proton irradiation
compared to approximately $65/\%$ kill rate by helium-ions. With
the pulse of 12 mGy/step at 15 keV/$\mu$m, we observed that both
protons and carbon-ions have similar cell killing effects,
however, there was a noticeable change of mutation frequency.

A relation between representative cell survival and dose-LET was
extracted after evaluating dose-LET nomenclature with light and
heavy ions. The surface plots of the representative survival curves are not displayed here. It is observed 
that as the average LET increases, the curves become more steeper and the survival curves with protons
converge faster than those from helium ions at low-LET.
For most cell types, survival curves start with a moderate slope,
and with increasing dose, the slope correspondingly increases. At
low doses, survival curves start with a steep slope which
decreases at a moderate dose and flattens out with increasing
dose. Therefore, the efficiency per dose increment decreases as
well. This can be understood in terms of the reparability of
radiation induced damages. At low doses, only a few damages are
induced with a large spatial separation, and a considerable
fraction of these damages can be repaired correctly. In contrast,
at high doses, the density of damages increases, leading to an
interaction of damages and thus a reduced fraction of repairable
damages. The term ''interaction'' has to be understood here in the
most general sense. On the one hand, it can happen, e.g., that
actually two individual damages are combined to form a more
complex type of damage. On the other hand, two damages produced in
close vicinity can lead to conflicting or competing repair
processes, also reducing the fraction of repairable damage.

Fig. \ref{fig9} (a-c panels) collects and displays the survival
curves obtained from the simulation outcomes of cells irradiated
with protons, helium-ion and carbon-ions and compared to the
experimental data selected from \cite{Guan2015}. For the low LET,
protons seem to be more effective in cell kill than helium-ions.
Statistically, only $15\%$ of the cancer cells survive after
20th-time step compared to more than $50\%$ surviving cells with
helium-ions at the same LET value. We also notice that surviving
curves with helium-ion show a plateau(not shown here) at high
doses. This might be due to the shielding of a fraction of cells
as a result of the range of helium-ion at these energies not
exceeding the width of the cell, thus causing a plateau even at
relatively modest surviving fraction levels. Whereas we have not
been able to simulate with light and heavy ions that are exactly
matched for LET, nevertheless, we notice that at moderate LET,
protons and carbon ions are more effective in cell kill than
doubly-charged particles of similar LET. It is evident from
RBE-LET relationship that the data for $^{4}He$ indicate a maximum
RBE at around 90 - 100 keV/$\mu$m and flattens out at higher LET
values, this would explain why helium-ions are less effective than
protons with the same LET for cell survival. The increased
lethality of protons compared to helium ions in cell survival data
are therefore consistent with studies that place importance on the
extent to which ionizations are clustered at the nano-scale.
\begin{figure}[!ht]
\begin{center}
\begin{tabular}{ccccccccc}
\scalebox{0.33}{\includegraphics{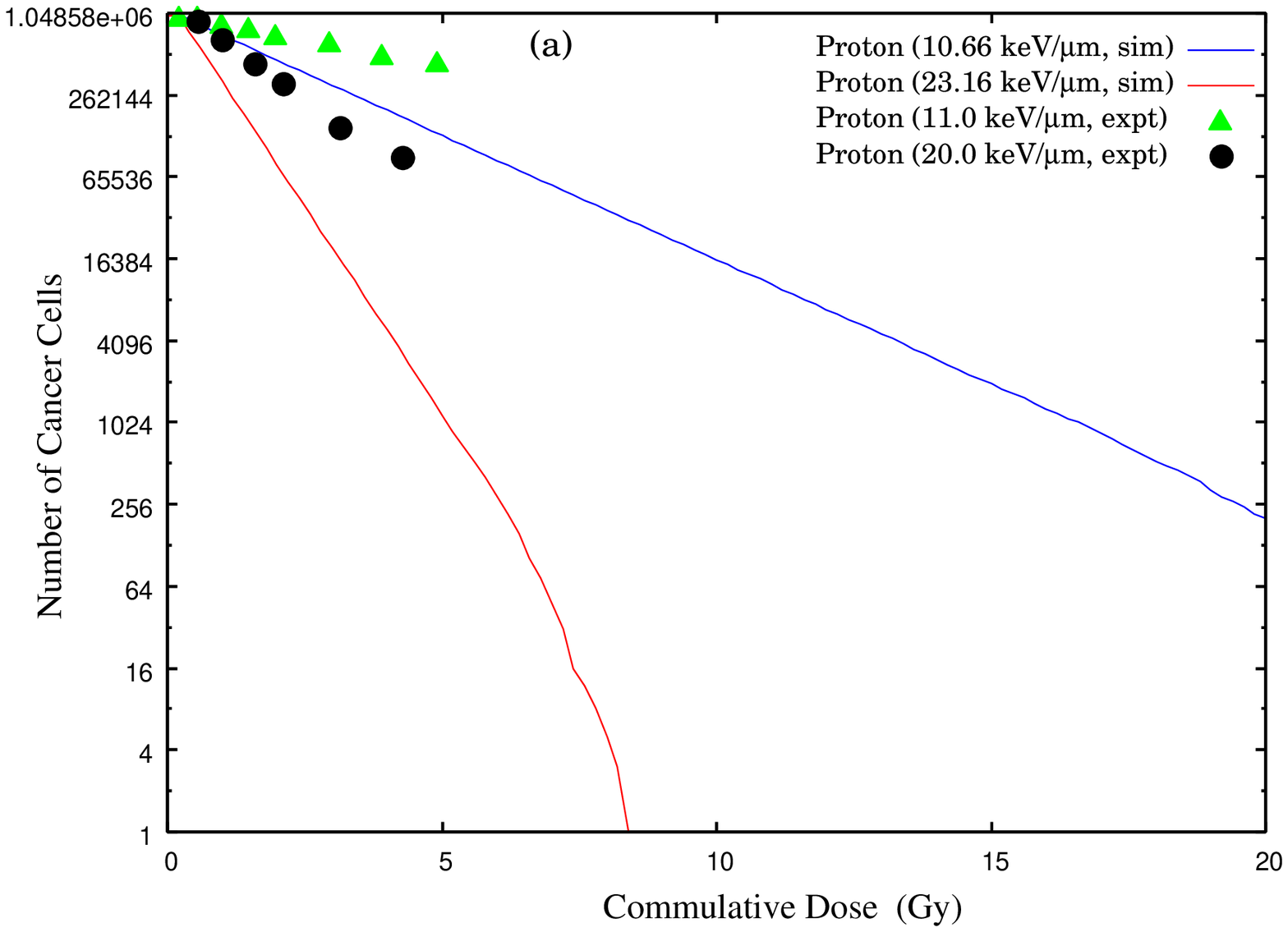}} &
\scalebox{0.33}{\includegraphics{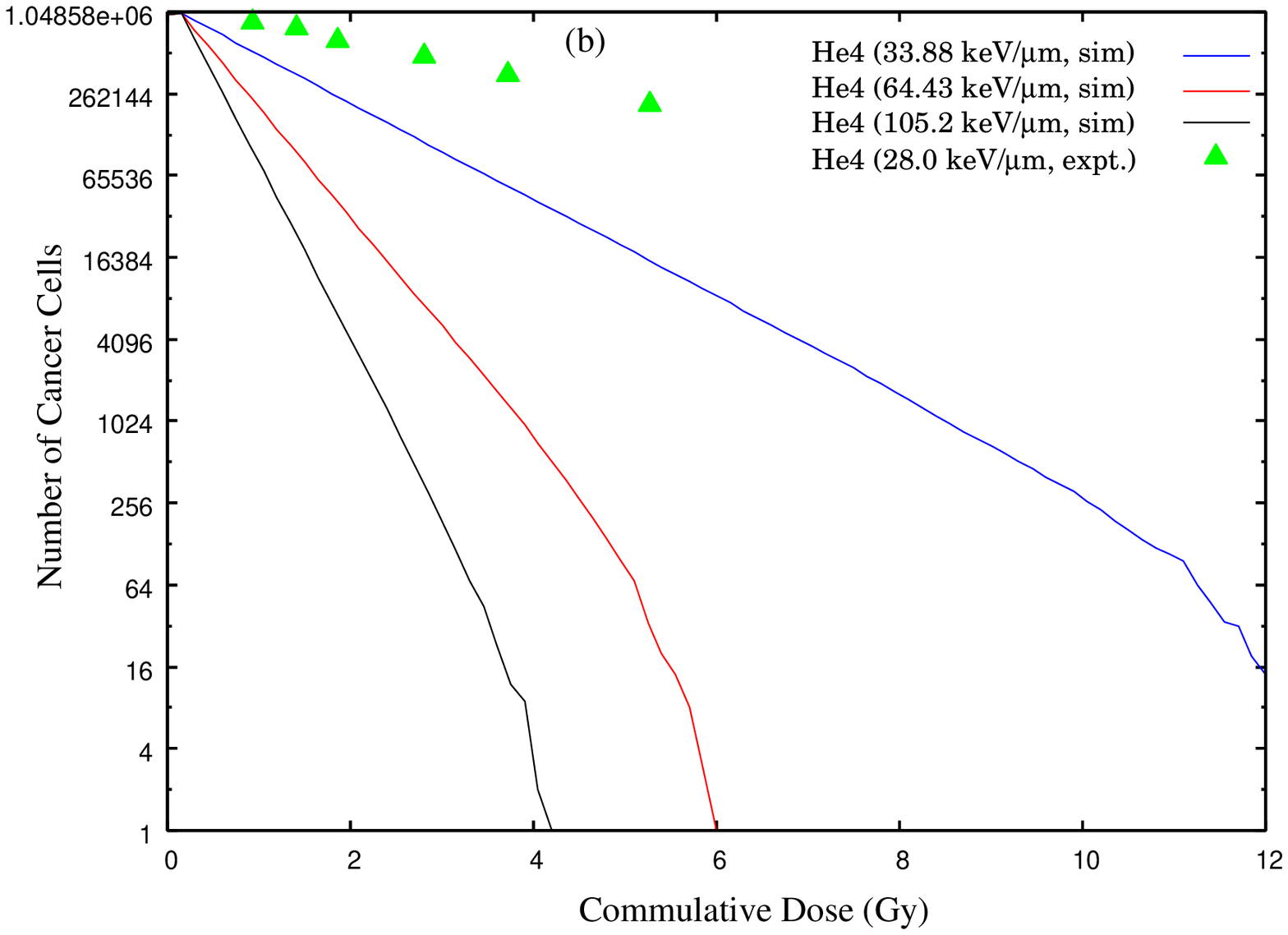}}\\
\scalebox{0.33}{\includegraphics{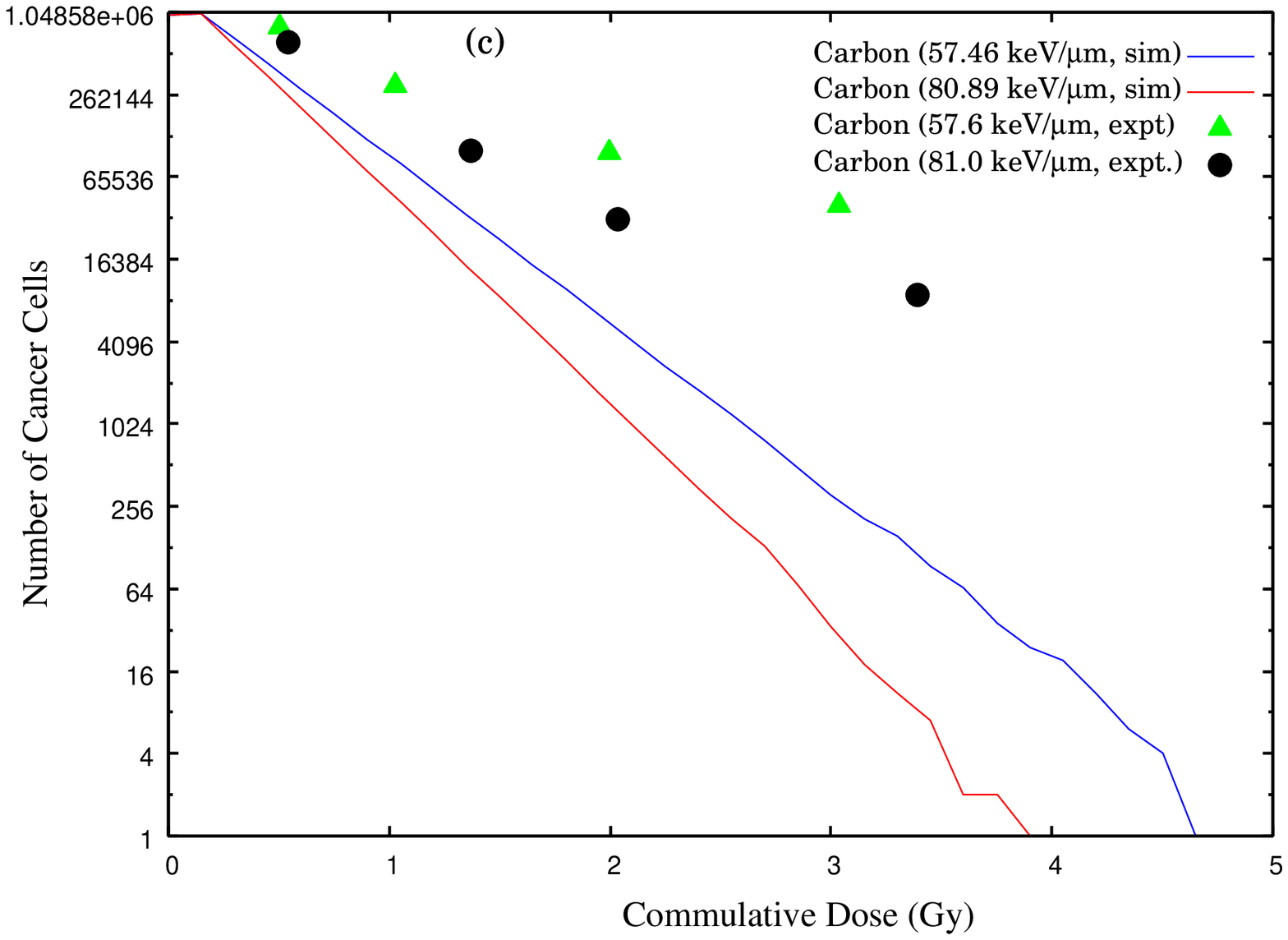}}&
\scalebox{0.33}{\includegraphics{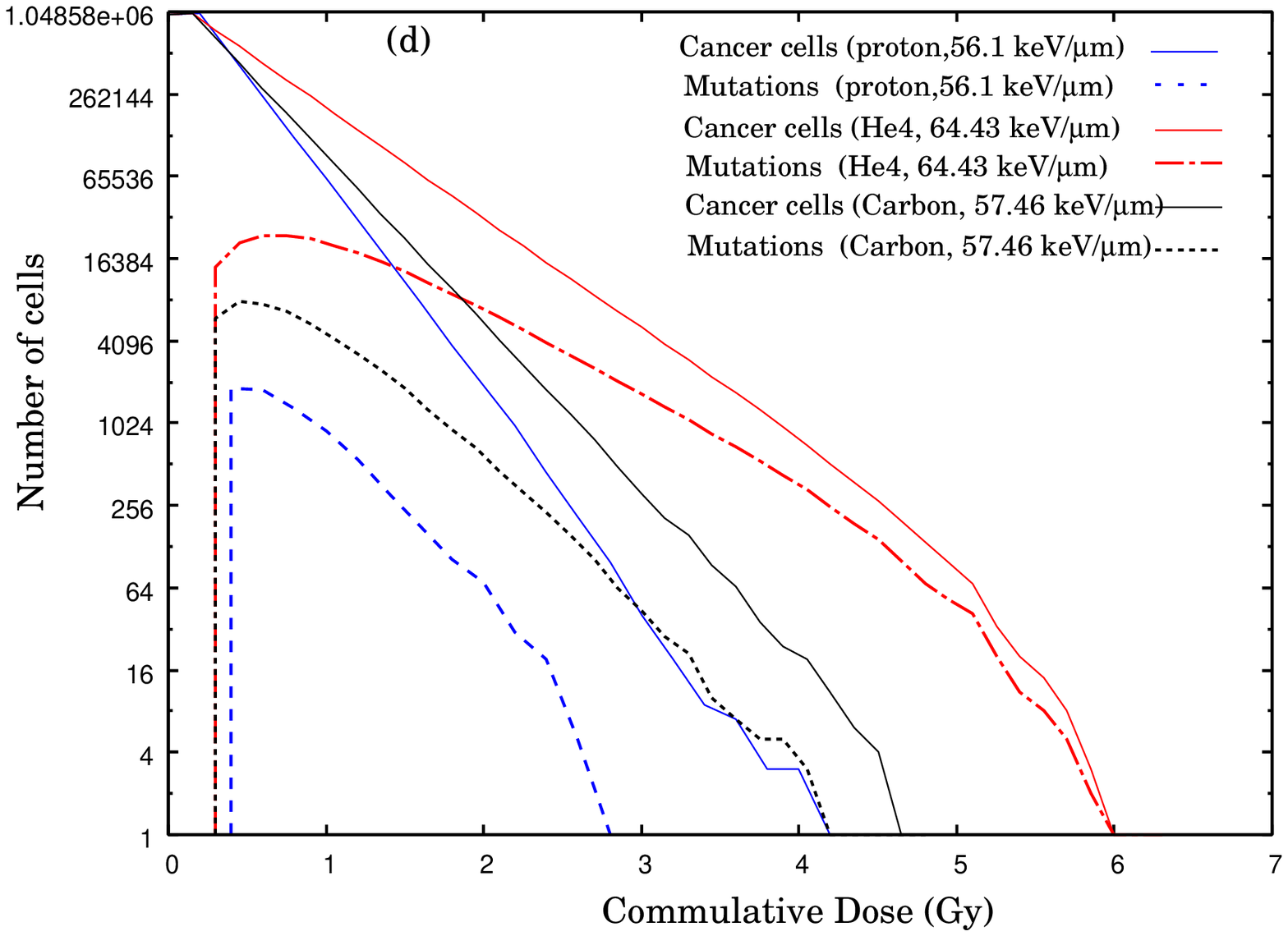}}
\end{tabular}
\caption{\label{fig9} Survival curves after exposure to protons
(panel a), helium ions (panel b), and carbon ions (panel c)
compared with the results from the experimental data
\cite{Guan2015}. Panel (d) shows the fraction of cancer and
mutation cells with different particle species.}
\end{center}
\end{figure}

In Fig. \ref{fig9} (b-d), we extend our investigation to survival
curves for cells exposed to helium-and carbon-ions both at medium
and high LET. Except for proton data at 56.1 keV/$\mu$m (panel d),
where we observed a low mutation frequency and a weak adaptive
response signal (not shown here for the sake of clarity), which
describes the ability of low-dose radiation to induce cellular
changes that alter the level of subsequent radiation-induced
damage, the survival curves start with a steep slope, and with
increasing dose, the slope decreases. Therefore, the efficiency
per dose increment decreases as well. We notice that the
dose-response relationship of the surviving curves is generally
non-linear. This is because of the quasi-linear/sigmoidal multiple
input dose-dependent relationships, internal conditions,
dependencies, and cancer cell transformation where simplified
biology was described by non-trivial probability distributions and
random processes that are close to real situations. Such a
behaviour of organism response has previously been observed in
many other studies \cite{Zyuzikov2011,Henriken2003}

The shape of the survival curves change from low dose in character
to high dose; the curves start showing a non-linear behaviour as
the cumulative dose increases. This might be related to the
reparability of radiation-induced damages and mutation frequency.
At high doses, only a few damages are induced due to shielding,
and a considerable fraction of these damages can be repaired
correctly. In contrast, at low and medium doses, the density of
damages increases, leading to an interaction of damages and thus a
reduced fraction of repairable damages. From the above surviving
curves it can be seen that at low and medium LET values here,
protons show increased effectiveness with increasing LET.

In light of the discussion above, it is reasonable to expect that
despite the discrepancies, our simulated survival curves for low
LET protons, helium- and carbon-ions are in good general agreement
with the considered experimental data. This serves one of the aims
of the model to predict the relationship between radiation-induced
DSBs in the cell nucleus and cell survival probability. This
supports the hypothesis of the mechanistic model that interaction
among DSBs induced by ionizing radiation contribute to the
quadratic term of the model, therefore the cell survival curves
are in agreement with the measured data for low-LET particle
species.

\section{Summary and Conclusion}

A stochastic Monte Carlo technique was used to simulate a full
grown colony of cancer cells using a mechanistic model of cellular
survival following radiation induced DSBs. Simulations were
performed using light and heavy-ion species over a range of LET
values. To make the model more realistic, we incorporated,
following the theories of nucleation and growth, individual
susceptibilities and probabilities as quasi-linear and sigmoidal
input relations described by various probability distributions and
random processes that apply for the effects like radiation-induced
mutation, cell transformation, and multiplication, hadroadaptive
response, cancer cell kill and more. This is a significant
improvement of the linear assumption for stochastic cancer effect
used frequently. Hypoxia was implemented through random assignment
of partial oxygen pressure values to individual cells during
tumour irradiation. The uncertainties in the measurements are
estimated by binning the numerical data into 10 blocks. The mean
and the final errors of the observables obtained using a
single-elimination jackknife method with each bin regarded as an
independent data point.

The yield of radiation induced DSB and DSB yield per cell per
primary particle was calculated using a fast Monte Carlo damage
simulation algorithm. At high LET, the estimated DSB yields showed
ion-specific characteristics. Above 12 keV/$\mu$m, proton showing
most effectiveness in inducing DSB than heavier ions at the same
LET. Our DSB yield results at low LET showed consistency with
those obtained using LEM model-based data with proton irradiation.

The LET-dependent DSB yield for different particle species was
used to calculate the $\alpha$, $\beta$ and $\alpha /\beta$ ratio,
using the improved descriptions of their definitions in the model.
The simulation results predict a gradual increase in $\alpha$,
$\beta$, and the ratio $\alpha /\beta$ as a function of LET which
is consistent qualitatively with the cell-inactivation target
theory. The ratio showed a quick increase with LET, indicating a
cluster DNA damage effect and a decrease in interaction of DSBs
induced by different primary particles. At high LET, the
contribution of quadratic parameter $\beta$ was found to be
vanishingly small. It was also observed that the cells with a
higher $\alpha /\beta$ ratio provide large RBEs for low and medium
doses compared with cells with smaller $\alpha /\beta$ ratios.
This implies that the tumours with high repair indicators
surrounded by healthy tissue with smaller repair indicators are
suited for hypofractionated regimens than from normal fraction
schemes and vice versa. At the medium doses, the RBE values seem
more or less insensitive to the ratio. This clearly indicates that
both LET and $\alpha /\beta$ ratio need to be taken as RBE
predictors for any hadrontherapy treatment plan. The predicted
estimates of RBE at the initial slopes and at $10\%$ survival that
are in good agreement with the experimental data. Since the values
of $\alpha$ and $\beta$ change with LET, the dependence of RBE on
particle species and the cellular repair capacity was modelled by
LET-dependent RBE model. These findings suggest that it is worth
considering at least main RBE dependencies in treatment planning,
and being particularly cautious for tissues with a low $\alpha
/\beta$ ratio. This is in contrast with some earlier studies which
support the conclusions that RBE dependence on cell type and
particle species is small enough to be safely neglected.

Finally, from the surviving curves, it becomes clear that although
most of the dose-LET and given observable relationships are
linear, the dose-response outcome is more complex than the one
predicted by the oversimplified LQ model. Another interesting
feature was the rapid increase in the probability of cell death
relative to probability of its multiplication and number of
mutations per cancer cell with increasing dose-LET values. Despite
this, the model still reproduces the behaviour of cell lines well
across a range of conditions without cell-line specific fitting
parameters. Thus while exact quantification of, for example,
$\alpha /\beta$ ratios may prove challenging for a specific
experimental condition, the model still has the potential to make
useful predictions about overall sensitivity.

While the current approach is sufficient to demonstrate the
viability of the model, explicitly incorporating models of the
underlying genetic pathways driving these effects will enable more
granular models of the impact of tumour genetics. The cell
survival probability does not reflect and explicate damage caused
by DNA single-strand breaks which could be many times more than
the number of DSBs, as well as SSB damage conversion into DSB
damage resulting in the stop of cell proliferation. This might
demand for an improved description of the biological and physical
process in cell survival probability defined LQ-theory. We intend
to improve the model and the algorithm to incorporate
cancer-specific factors, such as DNS single-strand break and cell
proliferation before reproductive death in the conventional
LQ-theory as well as complex biology of the cancer cells and more
complex tissue reactions.

\section{Acknowledgments}
We wish to thank Krzysztof Fornalski for a number of valuable
suggestions, which provided the impetus for much of this work. The
computations for DNA damage were done using a modified version of
publicly available MCDS code (see
http://faculty.washington.edu/trawets/mcds). We are also grateful
for access to KCST computing facility.

\section*{References}


\end{document}